\documentclass[acp, hvmath]{copernicus}

\begin{document}\hack{\sloppy}

\title{Comparison of VLT/X-shooter \chem{OH} and \chem{O_2} rotational temperatures with consideration of TIMED/SABER emission and temperature profiles}

\Author[1]{S.}{Noll}
\Author[2,1]{W.}{Kausch}
\Author[3,1]{S.}{Kimeswenger}
\Author[1]{S.}{Unterguggenberger}
\Author[4]{A.~M.}{Jones}

\affil[1]{Institute for Astro- and Particle Physics, University of Innsbruck, Technikerstr. 25/8, 6020 Innsbruck, Austria}
\affil[2]{Department of Astrophysics, University of Vienna, T\"urkenschanzstrasse 17, 1180 Vienna, Austria}
\affil[3]{Instituto de Astronom\'ia, Universidad Cat\'olica del Norte, Avenida Angamos 0610, Antofagasta, Chile}
\affil[4]{Max Planck Institute for Astrophysics, Karl-Schwarzschild-Str. 1, 85748 Garching, Germany}

\correspondence{S.~Noll (stefan.noll@uibk.ac.at)}

\runningtitle{Comparison of rotational temperatures}
\runningauthor{S. Noll et al.}

\received{23~September~2015}
\accepted{25~October~2015}
\published{}

\firstpage{1}

\maketitle

\begin{abstract}
  Rotational temperatures $T_{\text{rot}}$ derived from lines of the
  same \chem{OH} band are an important method to study the dynamics
  and long-term trends in the mesopause region near 87\,\unit{km}. To
  measure realistic temperatures, the rotational level populations
  have to be in local thermodynamic equilibrium (LTE). However, this
  might not be fulfilled, especially at high emission altitudes. In
  order to quantify possible non-LTE contributions to the \chem{OH}
  $T_{\text{rot}}$ as a function of the upper vibrational level
  $v^{\prime}$, we studied a sample of 343 echelle spectra taken with
  the X-shooter spectrograph at the Very Large Telescope at Cerro
  Paranal in Chile. These data allowed us to analyse 25 \chem{OH}
  bands in each spectrum. Moreover, we could measure lines of
  \chem{O_2}$b${\text{(0-1)}}, which peaks at about 94 to
  95\,\unit{km}, and \chem{O_2}$a${\text{(0-0)}} with an emission peak
  at about 90\,\unit{km}. The latter altitude is reached in the second
  half of the night after a rise of several \unit{km} because of the
  decay of a daytime population of excited \chem{O_2}. Since the
  radiative lifetimes for the upper levels of the two \chem{O_2} bands
  are relatively long, the derived $T_{\text{rot}}$ are not
  significantly affected by non-LTE contributions. These bands are
  well suited for a comparison with \chem{OH} if the differences in
  the emission profiles are corrected. For different sample averages,
  we made these corrections by using \chem{OH} emission,
  \chem{O_2}$a${\text{(0-0)}} emission, and \chem{CO_2}-based
  temperature profile data from the multi-channel radiometer SABER on
  the TIMED satellite. The procedure relies on differences of
  profile-weighted SABER temperatures. For
  an \chem{O_2}$a${\text{(0-0)}}-based reference profile at
  90\,\unit{km}, we found a good agreement of the \chem{O_2} with the
  SABER-related temperatures, whereas the \chem{OH} temperatures,
  especially for the high and even $v^{\prime}$, showed significant
  excesses with a maximum of more than 10\,\unit{K} for $v^{\prime} =
  8$. The exact value depends on the selected lines and molecular
  parameters. We could also find a nocturnal trend towards higher
  non-LTE effects, particularly for high $v^{\prime}$. The amplitude
  of these variations can be about 2\,\unit{K} or less, which tends to
  be significantly smaller than the total amount of the non-LTE
  contributions. The variations revealed may be important for
  dynamical studies based on $T_{\text{rot}}$ derived from \chem{OH}
  bands with high $v^{\prime}$.
\end{abstract}

\introduction
\label{sec:intro}

Studies of long-term trends as well as short-term variations of the
temperature in the mesopause region are often based on so-called
rotational temperatures $T_{\text{rot}}$ derived from emission lines
of the hydroxyl (\chem{OH}) airglow
\citep[e.g.][]{BEI03,BEI08,KHO08,REI14}, which is mostly caused by the
chemical reaction of hydrogen (\chem{H}) and ozone (\chem{O_3})
\citep{BAT50} at a~typical height of 87\,\unit{km}
\citep[e.g.][]{BAK88}. Ground-based instruments usually measure the
intensities of a~few lines with different rotational upper levels
$N^{\prime}$ of a~single band \citep[e.g.][]{BEI03,SCHM13}. Assuming
a~Boltzmann distribution of the populations at the different
rotational levels, a~temperature can be derived, which is reliable if
the frequency of thermalising collisions is high compared to the
natural radiative lifetime of a~state. The latter ranges from about 58
to 5\,\unit{ms} for the relevant upper vibrational levels $v^{\prime}
= 1$ to 9 of the \chem{OH} electronic ground state \citep{XU12}. For
the \chem{OH} peak emission height, this appears to be sufficiently
long to establish the assumed Boltzmann distribution \citep{KHO08}.

However, at higher altitudes with lower air density, collisions may be
too infrequent, especially for the higher $v^{\prime}$. Hence, the
measured $T_{\text{rot}}$ can deviate from the true temperatures due
to the contribution of a~population of \chem{OH} molecules, where the
condition of a~local thermodynamic equilibrium (LTE) for the
rotational levels is not fulfilled. Comparisons of $T_{\text{rot}}$
from various bands differing in $v^{\prime}$ show clear signatures of
such non-LTE effects \citep{COS07,NOL15}. The measured
$T_{\text{rot}}$ tend to increase with $v^{\prime}$ and even
$v^{\prime}$ appear to indicate higher $T_{\text{rot}}$ than the
adjacent odd $v^{\prime}$. For this reason, the highest
$T_{\text{rot}}$ are found for \chem{OH} bands with $v^{\prime} =
8$. As discussed by \citet{NOL15}, this can only be explained by
a~significant impact of a~nearly nascent vibrational and rotational
level distribution related to the hydrogen--ozone reaction, which
mostly populates $v^{\prime} = 9$ and 8 \citep{CHAR71,OHO85,ADL97}.
The differences in the results for even and odd $v^{\prime}$ can then 
be understood by the higher initial $T_{\text{rot}}$ for $v^{\prime} = 8$
\citep{LLE78} and the high probability of radiative transitions with 
$\Delta v = 2$ \citep{MIE74,ROU00,KHO08}. The $T_{\text{rot}}$ non-LTE 
contributions also appear to vary with time, as observed by 
\citet{NOL15}. At Cerro Paranal (24.6{\degree}\,S), they tend to 
enhance during the night when data of several years are averaged. This 
is probably caused by a~rising \chem{OH} emission layer, which can be 
inferred from a~decrease of the measured \chem{OH} intensity
\citep{YEE97,MEL99,LIU06,SAV15}.  Apart from $v^{\prime}$, the
measured $T_{\text{rot}}$ are also strongly affected by the choice of
lines. Higher $N^{\prime}$ tend to show stronger non-LTE contributions
\citep{PEN89,PEN93,PER92,COS07}. For this reason, only lines of the
first three or four rotational levels are usually used for
$T_{\text{rot}}$ derivations
\citep[e.g.][]{BEI03,PER07,COS07,SCHM13}. However, even this
restriction is not sufficient to obtain line-independent
$T_{\text{rot}}$. The \chem{OH} electronic ground level has two
substates, $X^2{\Pi}_{3/2}$ and $X^2{\Pi}_{1/2}$. Population
differences of several per cent were found by \citet{NOL15} for
rotational levels of the same substate and for comparisons between the
two substates.

The approach to estimate temperatures in the mesopause region by means
of airglow lines with different rotational upper levels can also be
applied to bands of molecular oxygen (\chem{O_2}). However, long
ground-based time series are only available for
\chem{O_2}($b^1\Sigma^{+}_{\mathrm{g}}-X^3\Sigma^{-}_{\mathrm{g}}$){\text{(0-1)}}
at about $0.865$\,\unit{\mu m} \citep[e.g.][]{KHO08,REI14}, which is
sufficiently bright, hardly blended by \chem{OH} lines, and not
affected by self-absorption in the lower atmosphere like the stronger
{\text{(0-0)}} band. Since \chem{O_2}$b${\text{(0-1)}} remains
unresolved with the standard instruments in use, a~simulation of the
temperature dependence of the band shape is required
\citep[e.g.][]{SCHE01,SHI07,KHO08}. In contrast to \chem{OH}, the
radiative lifetime $\tau$ of \chem{O_2}$b${\text{(0-1)}} is relatively
long. Using the Einstein-$A$ coefficients from the HITRAN2012
molecular line database \citep{ROT13}, $\tau$ is about 210\,s. This
suggests that \chem{O_2}$b${\text{(0-1)}} $T_{\text{rot}}$ are not
significantly affected by non-LTE contributions. The
$b^1\Sigma^{+}_{\mathrm{g}}$ state is essentially populated by atomic
oxygen recombination and subsequent collisions \citep{GRE81}, which
results in an emission profile with a~peak at about 94 to
95\,\unit{km} \citep[derived from the {\text{(0-0)}}
band;][]{WAT81,GRE86,YEE97}, i.e. \chem{O_2}$b${\text{(0-1)}} is
sensitive to a~different part of the mesopause region compared to
\chem{OH}. With full widths at half maximum (FWHM) of 8 to
9\,\unit{km} \citep{BAK88,YEE97} for both kinds of emission, there is
only a~minor overlap of their profiles.

For a~comparison with \chem{OH}, transitions from the
$a^1\Delta_{\mathrm{g}}$ state of \chem{O_2} to the ground state
$X^3\Sigma^{-}_{\mathrm{g}}$ are more promising since the measured
nighttime emission peak heights of the {\text{(0-0)}} band at
1.27\,\unit{\mu m} are close to those of \chem{OH}
\citep{MCD87,LOP89,LIP13}. During most of the night, they are at about
90\,\unit{km}. At the beginning of the night, the effective emission
heights can be several \unit{km} lower. This behaviour is caused by
the contribution of a~daytime-related $a^1\Delta_{\mathrm{g}}$
excitation by ozone photolysis to the nighttime-related atomic oxygen
recombination \citep{VAL63,MCD87,LOP89,LOP92}. The former is relevant
for the night since the spontaneous radiative lifetime of
$a^1\Delta_{\mathrm{g}}$ is about 75\,min due to an~Einstein-$A$
coefficient of about $2.2\times10^{-4}$\,\unit{s^{-1}} for the
{\text{(0-0)}} band \citep{LAF98,NEW99,GAM01}. The daytime emission
peaks at about 50\,\unit{km} \citep{EVA68,HOW90} and effectively rises
in altitude after sunset up to the nighttime peak due to the faster
decay of the $a^1\Delta_{\mathrm{g}}$ population by collisional
quenching at lower altitudes. Although non-LTE effects are negligible,
only \citet{MUL95} studied $T_{\text{rot}}$ based on the bright
\chem{O_2}$a${\text{(0-0)}} band. Their sample consists of
ground-based observations with a~Fourier transform spectrometer at
Maynooth, Ireland (53.2{\degree}\,N), and covers 91 twilight periods
over 18 months. The $T_{\text{rot}}$ measurements required simulations
of the band for different ambient temperatures due to the strong line
blending at their resolution. Moreover, \chem{O_2}$a${\text{(0-0)}} is
strongly affected by self-absorption, which required careful radiative
transfer calculations. These efforts might explain the lack of
\chem{O_2}$a${\text{(0-0)}} $T_{\text{rot}}$ in the literature. As
an~alternative, \chem{O_2}$a${\text{(0-1)}} can be used. Although this
band is fairly weak and partly blended with strong \chem{OH} lines, it
was studied by \citet{PER13} using spectroscopic data taken at
Zvenigorod, Russia (55.7{\degree}\,N), from 2010 to 2011. The band
emission had to be modelled to retrieve $T_{\text{rot}}$.

In this paper, we will compare $T_{\text{rot}}$ derived from 25
\chem{OH} bands, \chem{O_2}$b${\text{(0-1)}}, and
\chem{O_2}$a${\text{(0-0)}} to quantify the non-LTE effects related to
the \chem{OH} measurements. For this purpose, we will use 343
observations from the X-shooter echelle spectrograph \citep{VER11} at
the Very Large Telescope (VLT) of the European Southern Observatory
(ESO) at Cerro Paranal in Chile (2635\,m, 24{\degree}38$^{\prime}$\,S,
70{\degree}24$^{\prime}$\,W). The \chem{OH} data were already
discussed by \citet{NOL15}. As X-shooter covers the wavelength range
from 0.3 to 2.5\,\unit{\mu m} with a~resolving power of at least
$\lambda/\Delta\lambda \approx 3300$, the different \chem{OH} and
\chem{O_2} bands can be studied in parallel and without the need of
band simulations for the $T_{\text{rot}}$ retrieval. The resolution is
sufficient to measure individual lines even in
\chem{O_2}$a${\text{(0-0)}}.

We will not directly compare the derived
$T_{\text{rot}}$ since the studied bands probe different parts of the
mesopause temperature profile due to the differences in the altitude
distributions of the emission. In particular, the significant changes
in the emission profile of \chem{O_2}$a${\text{(0-0)}}
\citep{LOP92,GAO11,LIP13} have to be corrected. For this reason, we
will use volume emission rate (VER) and temperature profiles from
satellite observations to derive temperature corrections depending on
the VER differences of the bands to be compared. For this purpose,
data of the ten-channel broad-band radiometer SABER \citep{RUS99} on
the TIMED satellite launched in 2001 are most suited. The limb sounder
measures profiles of the crucial \chem{O_2}$a${\text{(0-0)}} band and
two \chem{OH}-related channels at 1.64 and 2.06\,\unit{\mu m}, which
probe different upper vibrational levels $v^{\prime}$ \citep[$4/5$ and
$8/9$; see][]{BAK07}. The latter is important since the \chem{OH}
emission peak heights depend on $v^{\prime}$
\citep{ADL97,XU12,SAV12}. The temperature profiles are derived from
the SABER \chem{CO_2} channels under the consideration of non-LTE
effects \citep{MERT01,MERT04,REM08,REZ15}. TIMED has
a~slowly-precessing polar orbit, which allows a~24\,\unit{h} coverage
at low and mid-latitudes within about 60~days \citep{RUS99}. Since
this results in only two narrow local time intervals for a~given site
and day of year, our study will focus on average X-shooter and SABER
data for certain periods.

The paper is structured as follows. We will start with the description
of the X-shooter and SABER data used for this study
(Sect.~\ref{sec:data}). Then, we will describe line measurements and
$T_{\text{rot}}$ retrieval from the X-shooter data, the derivation of
SABER-based profiles suitable for the measured airglow bands, and the
correction of the corresponding $T_{\text{rot}}$ for the differences
in the VER profiles (Sect.~\ref{sec:analysis}). In
Sect.~\ref{sec:results}, we will discuss the resulting estimates of
the non-LTE effects related to the \chem{OH} $T_{\text{rot}}$ and
their change with nighttime and season. Finally, we will draw our
conclusions (Sect.~\ref{sec:conclusions}).

\section{Data set}\label{sec:data}

\subsection{X-shooter}\label{sec:xshooter}

The X-shooter instrument \citep{VER11} of the VLT is an echelle slit
spectrograph that covers the very wide wavelength range from 0.3 to
2.5\,\unit{\mu m} by three arms called UVB (0.30 to 0.59\,\unit{\mu
  m}), VIS (0.53 to 1.02\,\unit{\mu m}), and NIR (0.99 to
2.48\,\unit{\mu m}). This allows one to study various airglow bands
simultaneously, as demonstrated by \citet{NOL15}, who investigated 25
\chem{OH}($v^{\prime}$-$v^{\prime\prime}$) bands between 0.58
and 2.24\,\unit{\mu m} with $v^{\prime}$ from 2 to 9. Since X-shooter
is an instrument for observing astronomical targets, airglow studies
have to rely on archival, unprocessed spectra, which differ in terms
of line of sight, exposure time, and slit-dependent spectral
resolution. For this study, we have used the same data set that was
selected and processed by \citet{NOL15}. The extended analysis of
these data, therefore, focuses on the \chem{O_2}$b${\text{(0-1)}} and
\chem{O_2}$a${\text{(0-0)}} bands (see Sect.~\ref{sec:lines}).

\citet{NOL15} reduced archival data from October~2009 (start of the
archive) to March~2013 using version V2.0.0 of the ESO public pipeline
\citep{MOD10}. The resulting wavelength-calibrated two-dimensional
(2-D) spectra were then collapsed into a~1-D spectrum by applying
a~median along the spatial direction.  For the flux calibration,
response curves were created based on data of spectrophotometric
standard stars \citep{MOE14} corrected for molecular absorption
\citep{SME15,KAU15} and atmospheric extinction \citep{PAT11}. Due to
the instability of the flat-field calibration lamps for the correction
of the pixel-to-pixel sensitivity variations in the 2-D echelle
spectra, the fluxes of long wavelengths beyond 1.9\,\unit{\mu m}
including the \chem{OH} bands {\text{(8-6)}} and {\text{(9-7)}} are
relatively uncertain. For 77\,{{\%}} of the data, this could be
improved by means of spectra of standard stars with known spectral
shapes, i.e. Rayleigh--Jeans slopes.

From the processed data, a~sample of 343 VIS- and NIR-arm spectra was
selected.  The selection criteria were a~minimum exposure time of
3\,\unit{min}, a~maximum slit width of 1.5$^{\prime\prime}$
corresponding to $\lambda/\Delta\lambda \approx 5400$ for the VIS arm
and 3300 for the NIR arm, and a~difference in start, end, and length
of the exposure between VIS and NIR arm of less than
5\,\unit{min}. The latter was necessary since the observations in the
different X-shooter arms can be performed independently. Moreover,
21\,{{\%}} of the combined VIS- and NIR-arm spectra were rejected due
to unreliable $T_{\text{rot}}$ in at least one of the 25 investigated
bands. This can be caused by low data quality, an~erroneous data
reduction, or critical contaminations by residuals of the observed
astronomical object. 36 spectra of the final sample could only be used
up to 2.1\,\unit{\mu m}, which excludes \chem{OH}{\text{(9-7)}},
because a~so-called $K$-blocking filter was applied \citep{VER11}. The
343 selected spectra show a~good coverage of nighttimes and seasons
and, therefore, appear to be representative of Cerro Paranal. For more
details, see \citet{NOL15}.

\subsection{SABER}\label{sec:saber}

\begin{figure}[t]
\includegraphics[width=80mm,clip=true]{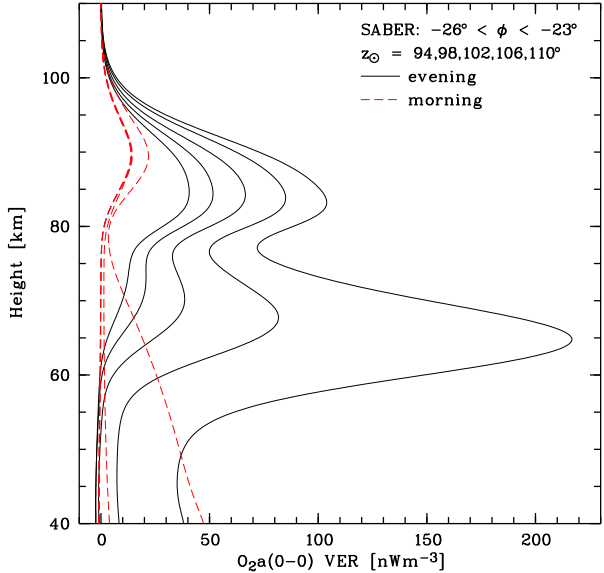}
\caption{Average SABER volume emission rate profiles of
  \chem{O_2}$a${\text{(0-0)}} during evening (solid lines) and morning
  twilight (dashed lines) for the latitude range from 23{\degree} to
  26{\degree}\,S and the years from 2009 to 2013. The plot shows
  profiles for five different solar zenith distances $z_\odot$ from
  94{\degree} to 110{\degree}. The VERs decrease with increasing
  $z_\odot$.}
\label{fig:VERO2a_twilight}
\end{figure}

\begin{figure}[t]
\includegraphics[width=80mm,clip=true]{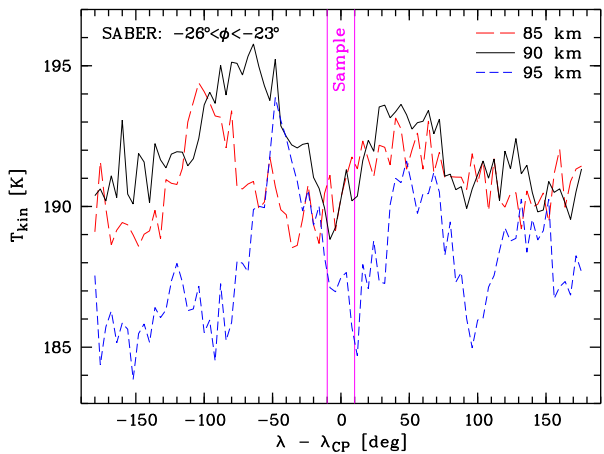}
\caption{SABER kinetic temperature $T_{\text{kin}}$ in \unit{K} as
  a~function of the longitude distance to Cerro Paranal $\lambda -
  \lambda_{\text{CP}}$ in degrees for the altitudes 85 (long dashed),
  90 (solid), and 95\,\unit{km} (short dashed) and bin sizes of
  4{\degree} and 1\,\unit{km}. The averages were derived from SABER
  data for the latitudes between 23{\degree} to 26{\degree}\,S, the
  years from 2009 to 2013, and zenith distances of at least
  100{\degree}. The final SABER sample selection criterion $|\lambda -
  \lambda_{\text{CP}}| \le 10${\degree} is visualised by vertical
  bars.}
\label{fig:Th_lon}
\end{figure}

Our analysis of VER and temperature profiles is based on version 2.0
of the limb sounding products of the SABER multi-channel photometer on
TIMED \citep{RUS99}. We used VER profiles from the
1.27\,\unit{\mu m} channel of \chem{O_2}$a${\text{(0-0)}} and the two
\chem{OH}-related channels centred on 1.64 and 2.06\,\unit{\mu m}. We
took the ``unfiltered'' data products, which are corrected VERs that 
also consider the emission of the targeted molecular band(s) outside 
the filter \citep{MLY05}. For \chem{O_2}$a${\text{(0-0)}}, the 
correction factor is 1.28 \citep{MLY07}. Finally, we used the SABER 
kinetic temperature $T_{\text{kin}}$ profiles. In version 2.0, they 
are based on the limb radiance measurements in the 4.3 and 
15\,\unit{\mu m} broad-band channels and a~complex retrieval algorithm 
involving non-LTE radiative transfer calculations \citep{REZ15}. For 
our analysis, we could use the VER and $T_{\text{kin}}$ profiles 
between 40 and 110\,\unit{km} because only for this altitude range all 
data are available. The lower and upper limits are related to the VERs 
and $T_{\text{kin}}$, respectively. We mapped all profiles (having 
a~vertical sampling of about 0.4\,\unit{km}) to a~regular grid with 
a~step size of 0.2\,\unit{km}.

To cover a~similar period as the X-shooter sample (see
Sect.~\ref{sec:xshooter}), we only considered SABER data of the years
2009 to 2013. The X-shooter data set is a~nighttime sample with
a~minimum solar zenith distance $z_\odot$ of 104{\degree}. For this
reason, we only selected SABER data with a~minimum $z_\odot$ of
100{\degree}. With this limit, it is guaranteed that the
\chem{O_2}$a${\text{(0-0)}} main emission peak is located in the
mesopause region, i.e. it is typical of nighttime conditions. The fast
change of the \chem{O_2}$a${\text{(0-0)}} VER profile at twilight is
illustrated in Fig.~\ref{fig:VERO2a_twilight}. As discussed in
Sect.~\ref{sec:intro}, the daytime population related to ozone
photolysis with a~maximum around the stratopause decays after
sunset. The late nighttime profile remains relatively constant until
the middle atmosphere is illuminated by the sun at dawn.

Since airglow emissions and temperatures significantly depend on
geographic location \citep[e.g.][]{YEE97,MEL99,HAY03,MAR06,XU10} due
to differences in the solar irradiation and activity of tides,
planetary and gravity waves, we selected only SABER data within the
range from 23{\degree} to 26{\degree}\,S. Cerro Paranal is located at
24.6{\degree}\,S. With this criterion and the exclusion of 14 profiles
which were suspicious in terms of the VER peak positions for the two
\chem{OH} channels, the resulting sample includes 27\,796
measurements. We further restricted the data set by also considering
longitudinal variations. Figure~\ref{fig:Th_lon} shows the SABER
$T_{\text{kin}}$ for the altitude levels 85, 90, and 95\,\unit{km} as
a~function of longitudinal distance $\lambda - \lambda_{\text{CP}}$ to
Cerro Paranal (70.4{\degree}\,W). On average, there are variations of
several \unit{K} and changes of the variability pattern depending on
altitude. For this reason, we limited the sample by the criterion
$|\lambda - \lambda_{\text{CP}}| \le 10${\degree}, which resulted in
1685 profiles. A~further reduction of the longitude range is not
prudent because then there could be too few profiles for
representative results for time and/or seasonal averages. We also
checked the results presented in the subsequent sections with respect
to the longitude limit (see Sect.~\ref{sec:tempcorrect}). Indeed, the
agreement of X-shooter and SABER variations does not significantly
improve with more restrictive limits. Also note that the SABER tangent
point can easily move by more than 100\,\unit{km} during a~profile
scan. Moreover, the TIMED yaw cycle of 60~days \citep{RUS99} causes
large gaps in the distribution of local times and days of year. For
odd months (e.g. March), the period around midnight is well covered at
Cerro Paranal. Even months show a~better evening/morning
coverage. Therefore, limitations in the correlation of X-shooter and
SABER data are unavoidable.

\section{Analysis}\label{sec:analysis}

This study mainly focuses on the quantification of non-LTE
contributions to \chem{OH} $T_{\text{rot}}$. The following section
describes the different steps of this analysis, which relies on
X-shooter-related \chem{OH} and \chem{O_2} $T_{\text{rot}}$
measurements and the SABER-related correction of temperature
differences by different band-specific emission profiles.

\subsection{Line intensities and rotational temperatures}\label{sec:lines}

The wavelength range, spectral resolution, and quality of the
X-shooter spectra (Sect.~\ref{sec:xshooter}) allowed us to derive
$T_{\text{rot}}$ from line intensities of 25 \chem{OH} bands,
\chem{O_2}$b${\text{(0-1)}}, and \chem{O_2}$a${\text{(0-0)}}.

\subsubsection{\chem{OH}}\label{sec:OH}

The $T_{\text{rot}}$ related to \chem{OH} are based on the line
selection and intensity measurements of \citet{NOL15}. The selection
only includes lines originating from rotational levels $N^{\prime} \le
4$ of both electronic substates $X^2{\Pi}_{3/2}$ and $X^2{\Pi}_{1/2}$
to minimise the non-LTE effects that increase with $N^{\prime}$
\citep{COS07,KHO08}. Moreover, it focuses on $P$-branch lines (related
to an~increase of the total angular momentum by 1) since these lines
are well separated in the X-shooter spectra. The \chem{OH} $\Lambda$
doublets \citep[e.g.][]{ROU00} are not resolved. Due to reasons like
blending by airglow lines of other bands or strong absorption in the
lower atmosphere (e.g. by water vapour), the number of suitable
$P$-branch lines varied between 2 and 8 for the 25 considered
\chem{OH} bands. In the case of \chem{OH}{\text{(6-4)}}, an additional
$R$-branch line was measured in order to have a~minimum of three lines
for quality checks. The measured intensities of all the lines were
corrected for molecular absorption in the lower atmosphere by means of
a~transmission curve for Cerro Paranal computed with the radiative
transfer code LBLRTM \citep{CLO05} by \citet{NOL12}. The absorption by
water vapour was adapted based on data for the precipitable water
vapour (PWV) retrieved from standard star spectra \citep{KAU15}. For
the \chem{OH} line shape, Doppler-broadened Gaussian profiles related
to a~temperature of 200\,\unit{K} were assumed.

The measurement of $T_{\text{rot}}$ is based on the assumption of
a~Boltzmann distribution of the $N^{\prime}$-related populations,
which are derived from $I/(g^{\prime}A)$, i.e. the measured intensity
$I$ divided by the $N^{\prime}$-dependent statistical weight 
$g^{\prime}$ (number of sublevels) times the Einstein-$A$ coefficient 
for the line. In this case, the natural logarithm of this population 
of a~hyperfine structure level of $N^{\prime}$, which we call $y$, is 
a~linear function of the level energy $E^{\prime}$. The line slope is 
then proportional to the reciprocal of $T_{\text{rot}}$ 
\citep{MEI50,MIE74}. \citet{NOL15} used the molecular parameters 
$E^{\prime}$, $A$, and $g^{\prime}$ from the HITRAN2012 database 
\citep{ROT13}, which are based on \citet{GOL98} for the investigated 
lines \citep{ROT09}, and from \citet{LOO07,LOO08}. We will mostly 
focus on the HITRAN-based parameters since they appear to provide 
more reliable $T_{\text{rot}}$.

\citet{NOL15} corrected the resulting $T_{\text{rot}}$ for the non-LTE
effects caused by the different band-dependent line sets (see
Sect.~\ref{sec:intro}) by introducing a~reference line set consisting 
of the three $P_1$- ($X^2{\Pi}_{3/2}$) and $P_2$-branch lines 
($X^2{\Pi}_{1/2}$) with $N^{\prime} \le 3$. The temperature changes
for the transition from the actual to the reference line set were
calculated based on the sample mean $T_{\text{rot}}$ of the 16 most
reliable bands. The resulting set-specific $\Delta T_{\text{rot}}$
were then applied to the temperatures for the individual observations.
For this study, we used a~different reference line set only including
the three $P_1$-branch lines. The corresponding results are easier to
interpret since the mixing of $P_1$- and $P_2$-branch lines causes
a~negative non-LTE effect, i.e. $T_{\text{rot}}$ decreases by the
lower relative population of the $X^2{\Pi}_{1/2}$ state, which has
higher $E^{\prime}$ than $X^2{\Pi}_{3/2}$. We calculated the
correction to the new reference set based on the four very reliable
\chem{OH} bands {\text{(4-1)}}, {\text{(5-2)}}, {\text{(6-2)}}, and
{\text{(7-4)}} \citep[cf.][]{NOL15}. This choice is not critical since
the slight $v^{\prime}$ dependence of the energy differences for the
rotational levels has only a~minor impact on the resulting $\Delta 
T_{\text{rot}}$. The $T_{\text{rot}}$ only based on the $P_1$ branch
are $3.0 \pm 0.3$\,\unit{K} higher than those given by \citet{NOL15}
if the molecular parameters are taken from HITRAN. In the case of
molecular data from \citet{LOO08}, $\Delta T_{\text{rot}}$ is $2.6 \pm
0.4$\,\unit{K}. If we focused on the fainter $P_2$-branch lines, the
corrections would be $4.3 \pm 1.0$\,\unit{K} and $5.9 \pm
2.5$\,\unit{K}, respectively.

Our study is based on mean $T_{\text{rot}}$ derived from bands with
the same $v^{\prime}$, which show very similar intensity and
$T_{\text{rot}}$ variations.  We took the $v^{\prime}$-specific
$T_{\text{rot}}$ from \citet{NOL15} (corrected for the different
reference line set) for each of the 343 sample spectra. Note that the
$v^{\prime} = 9$ mean values of 36 spectra with reduced wavelength
range do not include \chem{OH}{\text{(9-7)}} (see
Sect.~\ref{sec:xshooter}). The effect on the sample mean
$T_{\text{rot}}$ is negligible. The very faint \chem{OH}{\text{(8-2)}}
band is only considered by a~fixed $\Delta T_{\text{rot}}$ for
$v^{\prime} = 8$ derived from the sample mean
\citep[see][]{NOL15}. Excluding errors in the molecular parameters,
the uncertainties of the $T_{\text{rot}}$ ($v^{\prime}$) based on the
sample mean intensities are between 1.0\,\unit{K} for $v^{\prime} = 8$
and 2.0\,\unit{K} for $v^{\prime} = 2$. These values are mainly based
on the $T_{\text{rot}}$ variation for fixed $v^{\prime}$ and are
related to HITRAN data. For molecular data from \citet{LOO08}, the
corresponding uncertainties range from 1.8\,\unit{K} for $v^{\prime} =
9$ to 3.6\,\unit{K} for $v^{\prime} = 2$. The $T_{\text{rot}}$
uncertainties for individual observations are higher than the stated
values due to measurement errors depending on the signal-to-noise
ratio.

\subsubsection{$\chem{O_2}b{\text{(0-1)}}$}\label{sec:O2b}

\begin{figure}[t]
\includegraphics[width=80mm,clip=true]{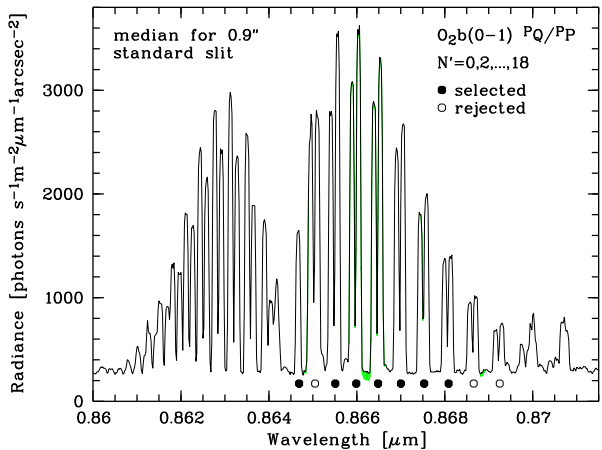}
\caption{Median X-shooter spectrum of \chem{O_2}$b${\text{(0-1)}} for
  the 189 VIS-arm sample spectra taken with the standard slit width of
  0.9$^{\prime\prime}$.  The original spectrum was corrected for solar
  absorption lines in the continuum (green areas). Pairs of $^{P}Q$
  and $^{P}P$ lines with even upper rotational levels $N^{\prime}$
  from 0 to 18 are marked by circles. Note that $^{P}Q$ (0) does not
  exist. The filled symbols indicate lines that were used for the
  derivation of $T_{\text{rot}}$.}
\label{fig:spec_O2b01}
\end{figure}

\begin{figure}[t]
\includegraphics[width=80mm,clip=true]{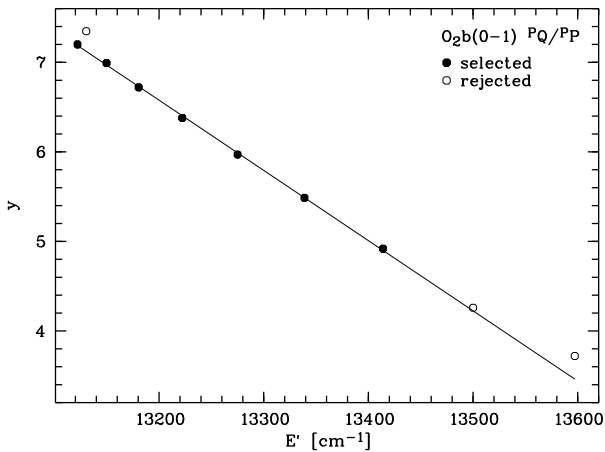}
\caption{Derivation of \chem{O_2}$b${\text{(0-1)}} $T_{\text{rot}}$
  under the assumption of a~Boltzmann distribution for the different
  rotational levels. The abscissa shows the HITRAN-based energy of
  a~given level in $\unit{cm}^{-1}$. The ordinate displays the natural
  logarithm of the HITRAN-based population of a~hyperfine-structure
  level relative to 1~Rayleigh second. Resulting from $^{P}Q$ and
  $^{P}P$ line intensities, the mean populations of the even
  rotational levels with $N^{\prime}$ from 0 to 18 for the full
  X-shooter sample are shown. Filled symbols mark reliable data points
  that were used for the derivation of $T_{\text{rot}}$ (see
  Fig.~\ref{fig:spec_O2b01}), which is illustrated by the solid
  regression line.}
\label{fig:y_Ep_O2b01}
\end{figure}

As discussed in Sect.~\ref{sec:intro}, \chem{O_2}$b${\text{(0-1)}} at
$0.865$\,\unit{\mu m} is the most popular non-\chem{OH} molecular band
for $T_{\text{rot}}$ retrievals in the mesopause region. It can also
be measured in the X-shooter spectra. With a~minimum resolving power 
of 5400 in the VIS arm, the individual lines of
\chem{O_2}$b${\text{(0-1)}} are not separated. However, in the $P$
branch, blended lines originate from the same upper rotational level
of $b^1\Sigma^{+}_{\mathrm{g}}$($v = 0$) (see
Fig.~\ref{fig:spec_O2b01}). This allows $T_{\text{rot}}$ to be
directly measured without the requirement for
$T_{\text{rot}}$-dependent synthetic spectra and data fitting, which
is the standard approach
\citep[e.g.][]{SCHE01,SHI07,KHO08}. Consequently, the systematic
uncertainties of the $T_{\text{rot}}$ measurement are expected to be
significantly lower. The two blended sub-branches are $^{P}Q$ and
$^{P}P$ \citep[e.g.][]{GAM98}, where the rotational level $N$
increases by 1 and the total angular momentum $J$ changes by 0 or $+1$
depending on the spin orientations in the lower state
$X^3\Sigma^{-}_{\mathrm{g}}$($v = 1$). Due to the symmetry of the
\chem{O_2} molecule, the upper state rotational levels $N^{\prime}$
have to be even. The selection rules also forbid $^{P}Q$ (0),
i.e. there is only a~single $P$-branch line for $N^{\prime} = 0$.

For the $T_{\text{rot}}$ determination, the intensities of the
$P$-branch line pairs with $N^{\prime}$ from 0 to 18 were measured
(see Fig.~\ref{fig:spec_O2b01}). The lines with $N^{\prime}$ higher
than 18 are already relatively faint and strongly blended with
\chem{OH} lines. Before the line intensities could be determined, the
continuum had to be measured and subtracted. Since an important part
of the continuum is caused by zodiacal light (sunlight scattered by
interplanetary dust particles) and/or scattered moonlight
\citep[e.g.][]{NOL12}, the strong solar \chem{Ca^{+}} absorption line
at 866.2\,\unit{nm} is critical for the continuum window between the
$N^{\prime} = 6$ and 8 lines (see Fig.~\ref{fig:spec_O2b01}). For this
reason, the solar flux atlas of \citet{WAL11} was used to calculate
a~continuum correction, which required the adaption of the solar
transmission to the corresponding resolution of the X-shooter spectra
and the determination of the variable fractional contribution of
a~solar-type spectrum to the continuum at
\chem{O_2}$b${\text{(0-1)}}. The latter was derived from the depth of
the \chem{Ca^{+}} absorption in the X-shooter data compared to the
solar spectrum and is important since there is also an~airglow
continuum related to reactions involving \chem{NO}
\citep{KHO08,NOL12,SEM14}. After the removal of the solar lines, the
median continuum level was measured in nine fixed windows, which were
optimised for the setups with the lowest spectral resolution,
i.e. a~slit width of 1.5$^{\prime\prime}$. The range between the
$N^{\prime} = 2$ and 6 lines could not be used for the continuum
determination due to significant contributions by \chem{OH} lines. The
resulting continuum fluxes were then linearly interpolated and
subtracted from the X-shooter spectra. Finally, the intensities of the
line pairs were integrated within limits which were adapted to the
spectral resolution. Since the $P$ branch of
\chem{O_2}$b${\text{(0-1)}} is not affected by molecular absorption
(cf. Sect.~\ref{sec:OH}), the corresponding correction could be
neglected. As the observations were carried out at different zenith
distances, the intensities were corrected to be representative of the
zenith by applying the equation of \citet{RHI21} \citep[see][]{NOL15}.

The resulting sample mean populations of the rotational levels
$N^{\prime} = 0$ to 18 are shown in Fig.~\ref{fig:y_Ep_O2b01}. Similar
to \chem{OH} described in Sect.~\ref{sec:OH} and \citet{NOL15}, they
correspond to $I/(g^{\prime}A)$, where $I$ and $A$ are sums for the
line pairs. The molecular parameters were taken from HITRAN2012
\citep{ROT13}, which is essentially based on \citet{GAM98} for
\chem{O_2}$b${\text{(0-1)}}. Most $y$ values show a~good correlation
with $E^{\prime}$. The outliers can be explained by contamination from
\chem{OH} lines. This result supports the assumption that non-LTE
effects are not relevant for the rotational level population. The
radiative lifetime of about 210\,\unit{s} derived from the HITRAN data
appears to be sufficiently long.

The optimal line set for the calculation of $T_{\text{rot}}$ turned
out to consist of $^{P}P$ (0) and the $P$-branch line pairs with
$N^{\prime}$ from 4 to 14. For the mean populations shown in
Fig.~\ref{fig:y_Ep_O2b01}, the corresponding systematic
$T_{\text{rot}}$ uncertainty is only 1.8\,\unit{K}. This line set
could be applied to 326 spectra. Due to instrumental issues or
contamination by astronomical objects, 15 fits were significantly
better by using only six instead of seven lines/line pairs. In two
cases, only five emission features were considered. The optimal line
sets were determined by an automatic iterative procedure. In order to
avoid systematic differences depending on the line set, the
$T_{\text{rot}}$ based on less than seven data points were
corrected. For this, the reliable spectra were used to derive the
differences between the $T_{\text{rot}}$ for the full and each reduced
line set.  The mean amount of the correction was 0.5\,\unit{K}.

We applied further corrections to the measured
$T_{\text{rot}}$. First, we investigated possible systematic errors
related to the widths of our continuum windows, which also had to be
suitable for the low-resolution data. For 23 reliable high-resolution
spectra taken with the 0.7$^{\prime\prime}$ slit, we could
significantly increase the number of continuum pixels. The resulting
temperatures were about 0.4\,\unit{K} lower than those obtained based
on the standard windows. As a~consequence, we decreased the
$T_{\text{rot}}$ of the whole sample by the same amount. Second, we
corrected the influence of the slit width on $T_{\text{rot}}$ by
adapting the resolution of the 23 0.7$^{\prime\prime}$ spectra to
those taken with a~0.9$^{\prime\prime}$, 1.2$^{\prime\prime}$, or
1.5$^{\prime\prime}$ slit and deriving the $T_{\text{rot}}$
differences. This resulted in temperature increases of about 0.2, 0.4,
and 1.0\,\unit{K}, respectively. All these corrections are relatively
small, which makes us confident that our $T_{\text{rot}}$ for
\chem{O_2}$b${\text{(0-1)}} are robust.

\subsubsection{$\chem{O_2}a{\text{(0-0)}}$}\label{sec:O2a}

\begin{figure}[t]
\includegraphics[width=80mm,clip=true]{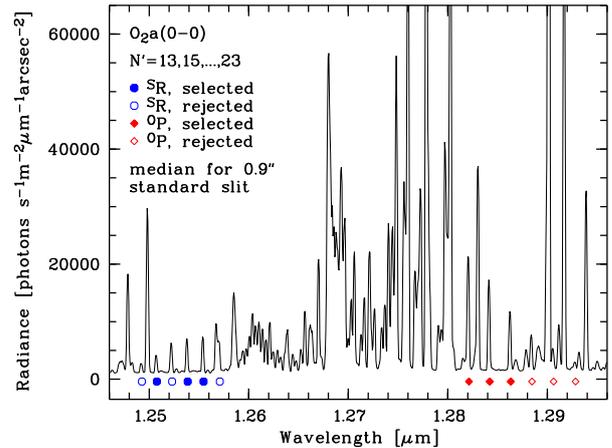}
\caption{Median X-shooter spectrum of \chem{O_2}$a${\text{(0-0)}} for
  the 177 NIR-arm sample spectra taken with the standard slit width of
  0.9$^{\prime\prime}$.  $^{S}R$ and $^{O}P$ lines with odd upper
  rotational levels $N^{\prime}$ from 13 to 23 are marked by circles
  and diamonds, respectively. The filled symbols indicate lines that
  were used for the derivation of $T_{\text{rot}}$.}
\label{fig:spec_O2a00}
\end{figure}

\begin{figure}[t]
\includegraphics[width=80mm,clip=true]{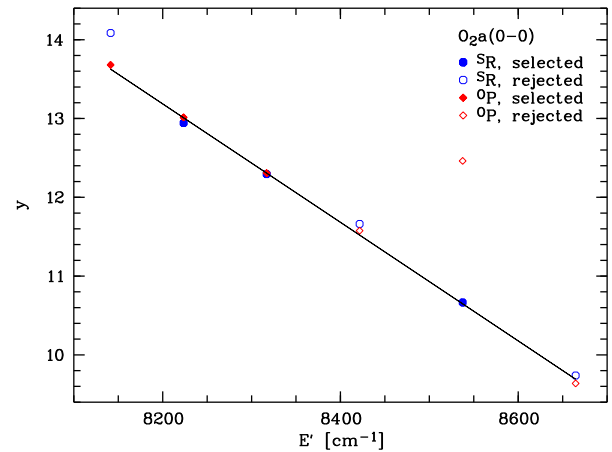}
\caption{Derivation of \chem{O_2}$a${\text{(0-0)}}
  $T_{\text{rot}}$. The plot is similar to
  Fig.~\ref{fig:y_Ep_O2b01}. Resulting from $^{S}R$ (circles) and
  $^{O}P$ (diamonds) line intensities, the mean populations of the odd
  rotational levels with $N^{\prime}$ from 13 to 23 for the full
  X-shooter sample are shown. Filled symbols mark reliable data points
  that were used for the derivation of $T_{\text{rot}}$ (see
  Fig.~\ref{fig:spec_O2a00}), which is illustrated by the solid
  regression line.}
\label{fig:y_Ep_O2a00}
\end{figure}

\begin{figure}[t]
\includegraphics[width=80mm,clip=true]{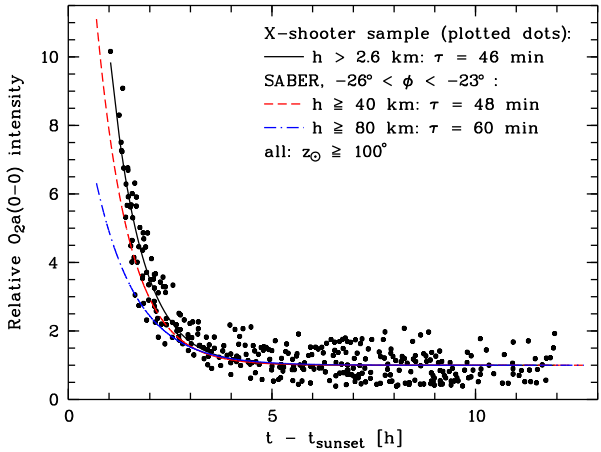}
\caption{\chem{O_2}$a${\text{(0-0)}} intensity relative to mean after
  midnight as a~function of the time since sunset for the
  X-shooter sample (dots). The solid line shows the best fit of
  a~constant and an~exponential function to the data. The plot also
  displays the corresponding fit results for the SABER sample without
  longitude restriction and lower intensity integration limits of 40
  (dashed) and 80\,\unit{km} (dash dotted). For the three fits, the
  resulting radiative lifetimes $\tau$ of the daytime population are
  shown in the legend.}
\label{fig:intO2a_time}
\end{figure}

The X-shooter NIR arm covers the strong \chem{O_2}$a${\text{(0-0)}}
band at 1.27\,\unit{\mu m} (see Fig.~\ref{fig:spec_O2a00}), which
contains nine branches related to the dominating magnetic dipole
transitions \citep[e.g.][]{LAF98,NEW99,ROU00,LES10}. With the minimum
resolving power of 3300 in the X-shooter NIR arm, the lines of two 
branches are sufficiently separated to allow individual intensity
measurements. This could not be performed by \citet{MUL95}, who have
done the only known \chem{O_2}$a${\text{(0-0)}} $T_{\text{rot}}$
measurements (see Sect.~\ref{sec:intro}). The suitable branches are
$^{S}R$ and $^{O}P$ (see Fig.~\ref{fig:spec_O2a00}), which have the
same upper levels but differ by a~change of $N$ by $-2$ and $+2$,
respectively. To fulfil the selection rules for $\Delta J$, these
$\Delta N$ have to be accompanied by an~opposite change of the spin
$\Delta S$ by 1. Only transitions with odd $N^{\prime}$ originate from
$a^1\Delta_{\mathrm{g}}$. For the line measurements, we focused on
intermediate $N^{\prime}$ between 13 and 23, which are sufficiently
strong and not too close to the highly blended branches in the band
centre.  The line intensities were measured in a~similar way as
described in Sect.~\ref{sec:O2b} for \chem{O_2}$b${\text{(0-1)}}. The
effect of solar lines on the continuum could be neglected. However,
since the lower state $X^3\Sigma^{-}_{\mathrm{g}}$($v = 0$) of
\chem{O_2}$a${\text{(0-0)}} is the ground state, the resulting line
intensities had to be corrected for the strong self-absorption at low
altitudes. The procedure was similar to the approach for \chem{OH}
described in Sect.~\ref{sec:OH} and \citet{NOL15}. Since the
transmission curves for Cerro Paranal with a~resolving power of about 
$1 \times 10^{6}$ calculated by \citet{NOL12} were not accurate enough
for \chem{O_2}$a${\text{(0-0)}}, we recalculated the \chem{O_2}
absorption with the maximum resolving power of $4 \times 10^{6}$ 
inherent to the radiative transfer code LBLRTM \citep{CLO05}. For 
a~nominal temperature of 190\,K for the \chem{O_2}$a${\text{(0-0)}} 
emission layer, we obtained zenith transmissions between 0.14 and 0.80 
for the measured $^{S}R$ lines and between 0.57 and 0.96 for the 
corresponding $^{O}P$ lines. The use of a~fixed temperature for the 
Doppler broadening calculation has only a~minor impact on the 
resulting $T_{\text{rot}}$. We performed a~second iteration of the 
transmission correction using the measured $T_{\text{rot}}$. The mean
$T_{\text{rot}}$ change compared to the first iteration was only about
$0.1$\,\unit{K}.

Figure~\ref{fig:y_Ep_O2a00} shows the sample mean
populations for the upper states of the measured
\chem{O_2}$a${\text{(0-0)}} lines. The populations were calculated by
means of the molecular parameters of the HITRAN2012 database
\citep{ROT13}, which are based on several studies
\citep{GAM98,NEW99,WAS06,LES10}. Most lines indicate a~tight linear
relation of $y$ and $E^{\prime}$. The good agreement of the
populations derived from $^{S}R$ and $^{O}P$ lines with the same
$N^{\prime}$ is also convincing. The few outliers can be explained by
blends with \chem{O_2} lines from other branches and \chem{OH}
lines. This result supports the expected good thermalisation of the
rotational level populations of $a^1\Delta_{\mathrm{g}}$($v = 0$) due
to the long radiative lifetime. The HITRAN data indicate
an~Einstein-$A$ coefficient of $2.28 \times 10^{-4}$\,\unit{s^{-1}}
(cf. Sect.~\ref{sec:intro}), which corresponds to $\tau \approx
73$\,\unit{min}.

We can directly constrain $\tau$ from our data. Since the daytime
production of $a^1\Delta_{\mathrm{g}}$ by ozone photolysis stops at
sunset, the subsequent decrease of the \chem{O_2}$a${\text{(0-0)}}
intensity is an~indicator of $\tau$
\citep[e.g.][]{MUL95,PEN96,NAI09}. Since there are also
$a^1\Delta_{\mathrm{g}}$ losses by collisions, the derived lifetime is
only a~lower limit. Figure~\ref{fig:intO2a_time} shows the
\chem{O_2}$a${\text{(0-0)}} intensities normalised to the mean after
midnight as a~function of the time after sunset for our X-shooter data
set. The steep intensity decrease at the beginning of the night is
evident. We fit this decrease by an exponential function plus 1. The
latter reflects the population related to the nighttime
$a^1\Delta_{\mathrm{g}}$ excitation, which is considered to be
constant. This assumption is reasonable since the average nighttime
variations are expected to be much weaker than those of the
daytime-related population. The resulting lower lifetime limit is
$46.1 \pm 1.6$\,\unit{min}, which is within the range of other
ground-based measurements \citep{MUL95,PEN96}. We can also derive
$\tau$ from our SABER data set (Sect.~\ref{sec:saber}). For the
latitude range from 23{\degree} to 26{\degree}\,S, $z_\odot$ greater
than 100{\degree}, and a~minimum height of 40\,\unit{km}, we obtained
about 48\,\unit{min}, which is close to the X-shooter value. The SABER
best-fit curve in Fig.~\ref{fig:intO2a_time} is shifted to earlier
times due to the sparse twilight sampling of the X-shooter data with
a~minimum $z_\odot$ of 105{\degree}. This could also contribute to the
small difference of the $\tau$ values. As illustrated by
Fig.~\ref{fig:VERO2a_twilight}, the decay of the daytime population is
faster at lower altitudes, where collisions are more
frequent. Therefore, the resulting $\tau$ depends on the considered
altitude range. For a~lower limit of 80\,\unit{km}, we obtained about
60\,\unit{min} (see Fig.~\ref{fig:intO2a_time}). Also based on SABER
data, \citet{NAI09} derived an~Einstein-$A$ coefficient consistent
with the theoretical value for a~height of 85\,\unit{km}.

As shown by Figs.~\ref{fig:spec_O2a00} and \ref{fig:y_Ep_O2a00}, the
optimal line set for the $T_{\text{rot}}$ retrieval consists of
$^{O}P$ (13), $^{S}R$ (15), $^{O}P$ (15), $^{S}R$ (17), $^{O}P$ (17),
and $^{S}R$ (21). In principle, the two lines with $N^{\prime} = 23$
could also have been included since the $T_{\text{rot}}$ derived from
the mean populations in Fig.~\ref{fig:y_Ep_O2a00} only differs by
$0.1$\,\unit{K} from the result for our standard line set. However,
they are relatively faint, which causes significant uncertainties if
the $T_{\text{rot}}$ are obtained from individual spectra. Using the
mean populations and the extended line set, the systematic
$T_{\text{rot}}$ uncertainty from the regression analysis is about
$2.1$\,\unit{K}.

The resulting $T_{\text{rot}}$ were corrected for different systematic
effects.  For this purpose, the 47 spectra taken with the narrow
0.4$^{\prime\prime}$ and 0.6$^{\prime\prime}$ slits were used. First,
we checked the quality of the crucial \chem{O_2} self-absorption
correction. This was achieved by multiplying the optical depth of all
lines by a~line-independent but variable factor and recalculating
$T_{\text{rot}}$ to find the results with the lowest regression
uncertainties. For the 47 test spectra, this procedure resulted in an
optimal factor of $1.03 \pm 0.03$ and a~$T_{\text{rot}}$ decrease by
$1.1 \pm 0.8$\,\unit{K}. This small correction confirms the reliability
of our radiative transfer calculations. Then, similar corrections as
for \chem{O_2}$b${\text{(0-1)}} (Sect.~\ref{sec:O2b}) were
performed. With continuum windows optimised for the narrow slits
instead of the wide 1.5$^{\prime\prime}$ slit, we obtained a~$\Delta
T_{\text{rot}}$ of $-0.3$\,\unit{K}. The effect of the spectral
resolution on the line measurements resulted in $T_{\text{rot}}$
corrections of $+1.2$, $+0.4$, and $+1.4$\,\unit{K} for the
0.9$^{\prime\prime}$, 1.2$^{\prime\prime}$, and 1.5$^{\prime\prime}$
slits. Taking the 0.6$^{\prime\prime}$ slit as a~reference, the
$T_{\text{rot}}$ of the three spectra obtained with
a~0.4$^{\prime\prime}$ slit were also corrected and increased by
3.5\,\unit{K}. This relatively large correction is probably related to
the starting resolution of underlying faint lines. In general, the
applied corrections are small and partly cancel each other out. The
systematic $T_{\text{rot}}$ uncertainty slightly increases from 2.1 to
2.2\,\unit{K} when the uncertainties of the $T_{\text{rot}}$
corrections are also considered.

\subsection{Emission profiles}\label{sec:profiles}

In order to compare the $T_{\text{rot}}$ derived from the discussed
bands (Sect.~\ref{sec:lines}), the differing height distributions of
the emission have to be considered since the measured bands probe
different parts of the mesopause temperature profile. Since the
ground-based X-shooter data do not provide this information, we
obtained VER and temperature profiles representative of Cerro Paranal
from the satellite-based SABER data \citep{RUS99}.

\subsubsection{\chem{OH}}\label{sec:OHprof}

As described in Sect.~\ref{sec:saber}, SABER has two \chem{OH}-related
photometric channels centred on 1.64 and 2.06\,\unit{\mu m} 
\citep{RUS99}, which essentially cover the {\text{(4-2)}}, 
{\text{(5-3)}} and {\text{(8-6)}}, {\text{(9-7)}} bands, respectively
\citep[see][]{BAK07}. The SABER \chem{OH} profiles cannot be used
directly since the \chem{OH} emission height depends on $v^{\prime}$
\citep{ADL97,XU12,SAV12}, which ranges from 2 to 9 for our X-shooter
data (Sect.~\ref{sec:OH}). For this reason, we had to derive
$v^{\prime}$-related profiles from the SABER data.

It is reasonable to assume that the $v^{\prime}$-dependent emission
peak altitudes are equidistant \citep[cf.][]{SAV12} (see also
Sect.~\ref{sec:correl}). Even if there are significant deviations from
this assumption, the maximum errors should only be of the order of
a~few 100\,\unit{m}, which is still relatively small compared to the
typical FWHM of 8 to 9\,\unit{km} of the \chem{OH} VER profiles
\citep[e.g.][]{BAK88}. Then, the $v^{\prime}$-dependent emission peak
altitudes can be derived by means of effective $v^{\prime}$ for the
two SABER channels. We achieved this by calculating \chem{OH} line
spectra from the vibrational level populations and $T_{\text{rot}}$
given by \citet{NOL15} for HITRAN data and convolving them with the
SABER filter curves \citep{BAK07}. In this way, we obtained fractional
contributions of the different \chem{OH} bands and, finally, the
effective $v^{\prime}$, which are 4.57 and 8.29 for the 1.64 and
2.06\,\unit{\mu m} channels, respectively. Although we neglected the
higher relative populations at high $N^{\prime}$ compared to those
related to $T_{\text{rot}}$ \citep[e.g.][]{COS07} and did not consider
the variability of the relative populations \citep{NOL15}, the
resulting values are sufficiently robust for our purpose. The
corresponding altitude uncertainty should be well below
100\,\unit{m}. From $\Delta v^{\prime} = 3.72$ and the emission peak
altitudes $h_{\text{peak}}$ of both \chem{OH} channels, we derived
a~mean altitude difference of 0.37\,\unit{km} for $\Delta v^{\prime} =
1$ and our Cerro Paranal SABER sample consisting of 1685 measurements
(Sect.~\ref{sec:saber}). This is in good agreement with data from
other studies \citep{SAV12}. Since $h_{\text{peak}}$ is not very
accurate due to a~step size of 0.2\,\unit{km} (Sect.~\ref{sec:saber}),
we used the centre of the range limited by the interpolated half
maximum positions $h_{\text{cen}}$ as a~more robust measure to
calculate the $v^{\prime}$-dependent profile altitudes. In this case,
the mean $\Delta v^{\prime} = 1$ altitude difference is
0.39\,\unit{km} with a~standard deviation $\sigma$ of 0.15\,\unit{km}.

\begin{figure}[t]
\includegraphics[width=80mm,clip=true]{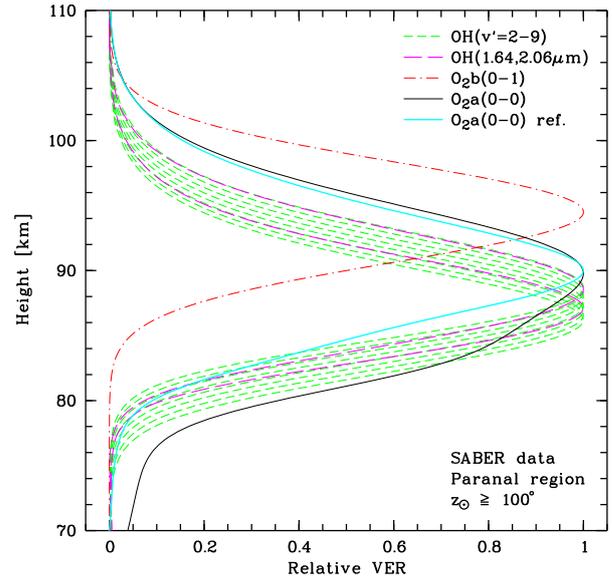}
\caption{Mean emission profiles of different molecular transitions
  (see legend) for the Cerro Paranal SABER sample. The volumn emission
  rates scaled to the emission peak are given as a~function of the
  height in \unit{km}. The emission peak height of the \chem{OH}
  profiles increases with $v^{\prime}$ and the mean profile of the
  2.06\,\unit{\mu m} SABER channel peaks higher than the
  1.64\,\unit{\mu m} one.}
\label{fig:rVER_all}
\end{figure}

The VER profiles of the two \chem{OH} channels (see
Fig.~\ref{fig:rVER_all}) show slightly different shapes. For the Cerro
Paranal SABER sample, the mean FWHM are 8.1 and 9.1\,\unit{km} for the
1.64 and 2.06\,\unit{\mu m} channels, respectively. The corrresponding
$\sigma$ are 1.9 and 2.1\,\unit{km}. For a~better comparability of the
$v^{\prime}$-dependent profiles and the related temperatures (see
Sect.~\ref{sec:tempcorrect}), we used merged profiles. For all SABER
measurements, we shifted both profiles normalised to the same
integrated intensity to the calculated $h_{\text{cen}}$ for each
$v^{\prime}$ and averaged them. The resulting mean profiles for the
Cerro Paranal SABER sample are shown in Fig.~\ref{fig:rVER_all}. They
have a~FWHM of 8.6\,\unit{km} with a~$\sigma$ of 1.9\,\unit{km}. The
average $h_{\text{cen}}$ range from 86.2\,\unit{km} ($v^{\prime} = 2$)
to 89.0\,\unit{km} ($v^{\prime} = 9$) with a~scatter between 1.3 and
1.4\,\unit{km}. The profiles are asymmetric. The mean $h_{\text{cen}}$
are about 0.6\,\unit{km} ($\sigma = 1.1$\,\unit{km}) higher than the
corresponding $h_{\text{peak}}$. For the effective height
$h_{\text{eff}}$, i.e.  the VER-weighted altitude, the average
difference is 1.2\,\unit{km}. The calculation of $h_{\text{eff}}$
excluded altitudes with negative VERs and $|h - h_{\text{cen}}| >
15$\,\unit{km} (see also Sect.~\ref{sec:tempcorrect}).

\subsubsection{$\chem{O_2}b{\text{(0-1)}}$}\label{sec:O2bprof}

SABER does not provide VER profiles for
\chem{O_2}$b${\text{(0-1)}}. For this reason, we assumed a~simple
Gaussian profile, which appears to be justified by the relatively
small measured asymmetries \citep[see][]{KHO08}. Based on values from
the literature (see Sect.~\ref{sec:intro}), where the measurements are
usually based on \chem{O_2}$b${\text{(0-0)}} \citep[e.g.][]{YEE97},
peak height and FWHM were set to 94.5 and 9\,km (see
Fig.~\ref{fig:rVER_all}).  Since our \chem{O_2}$b${\text{(0-1)}}
profile is fixed, we checked the influence of variations of
$h_{\text{peak}}$ and FWHM on the derived temperatures (see
Sect.~\ref{sec:tempcorrect}). For this analysis, we assumed the same
variations of these parameters as described in Sect.~\ref{sec:OHprof}
for \chem{OH}. As a~result, we found for individual profiles
an~average $\sigma$ of 2.0\,K. However, for the sample mean (where
transient waves vanish), $\sigma$ is only 0.5\,K. Since we are mainly
interested in average properties, the exact choice of
$h_{\text{peak}}$ and FWHM for \chem{O_2}$b${\text{(0-1)}} is not
critical for our study.

\subsubsection{$\chem{O_2}a{\text{(0-0)}}$}\label{sec:O2aprof}

\begin{figure}[t]
\includegraphics[width=80mm,clip=true]{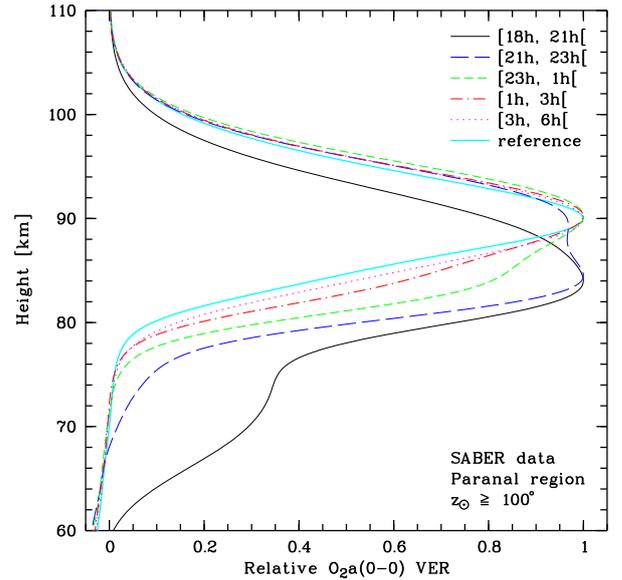}
\caption{\chem{O_2}$a${\text{(0-0)}} VER profiles scaled to the
  emission peak for the Cerro Paranal SABER sample. Mean profiles for
  five different nocturnal local time (LT) periods are shown (see
  legend). For comparison, the \chem{O_2}$a${\text{(0-0)}}-based
  reference profile is also displayed (light solid line).}
\label{fig:rVERO2a_nbina}
\end{figure}

SABER has a~dedicated channel for \chem{O_2}$a${\text{(0-0)}} (see
Sect.~\ref{sec:saber}). This was one of the reasons for selecting
SABER for this study since this band shows strong variations of the
emission profile. In Fig.~\ref{fig:VERO2a_twilight}, this was
illustrated for twilight conditions. As our X-shooter data set is
a~nighttime sample, we show the average emission profiles for the
Cerro Paranal SABER sample at five different time intervals in
Fig.~\ref{fig:rVERO2a_nbina}. For all periods, most of the emission
originates close to the \chem{OH} emission layer (see also
Fig.~\ref{fig:rVER_all}), which is crucial for a~reliable
$T_{\text{rot}}$ comparison (see
Sect.~\ref{sec:tempcorrect}). However, the emission peak indicates
a~clear transition from about 84 to 90\,\unit{km} in the first half of
the night. The early-night peak is obviously caused by a~daytime ozone
(and its photolysis) maximum in the upper mesosphere, whereas the
late-night peak appears to be related to atomic oxygen recombination
and subsequent collisions \citep[e.g.][]{LOP89,MUL95,KAU03} (see also
Sect.~\ref{sec:intro}).

\begin{figure}[t]
\includegraphics[width=80mm,clip=true]{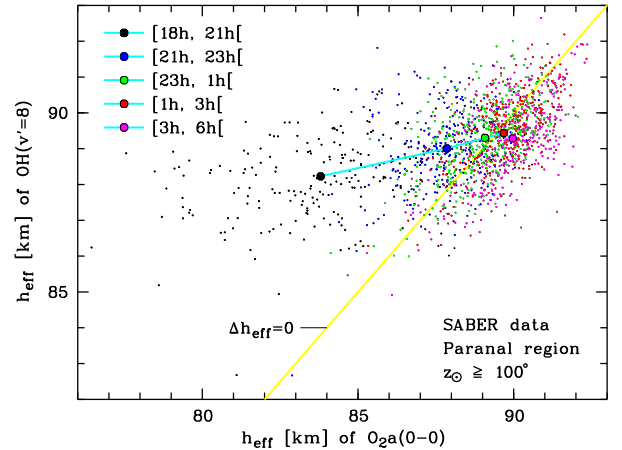}
\caption{Comparison of effective emission heights $h_{\text{eff}}$ in
  \unit{km} for \chem{OH}($v^{\prime} = 8$) and
  \chem{O_2}$a${\text{(0-0)}} for the Cerro Paranal SABER sample. The
  mean values for five different nighttime LT periods are shown by big
  circles of different colours connected by lines (see legend). The
  colours of the individual data points correspond to those of the
  related mean values.}
\label{fig:heff_nbina}
\end{figure}

The nocturnal variations of \chem{O_2}$a${\text{(0-0)}} are
illustrated in Fig.~\ref{fig:heff_nbina}. Here, the effective heights
$h_{\text{eff}}$ are given compared with those of
\chem{OH}($v^{\prime} = 8$). The calculation of the
\chem{O_2}$a${\text{(0-0)}} $h_{\text{eff}}$ excluded altitudes with
negative VERs, which essentially appear in the lower mesosphere (see
Fig.~\ref{fig:rVERO2a_nbina}). Additional height limits as for
\chem{OH} (see Sect.~\ref{sec:OHprof}) were not applied because of the
strong profile variations. The general profile limits of 40 and
110\,\unit{km} (Sect.~\ref{sec:saber}) are not an~issue for the
nighttime. Figure~\ref{fig:heff_nbina} shows a~relatively large
scatter of 2.4\,\unit{km} in the \chem{O_2}$a${\text{(0-0)}}
$h_{\text{eff}}$ at the beginning of the night, which is not seen for
\chem{OH}($v^{\prime} = 8$) ($\sigma = 1.1$\,\unit{km}). In the course
of the night, the deviation (4.4\,\unit{km}) and scatter become
smaller until a~convincing linear relation with \chem{OH}($v^{\prime}
= 8$) is established. The correlation coefficient for the last period
is $r = 0.62$, which is significantly higher than $r = 0.35$ for the
first period. The increase of $h_{\text{eff}}$ for
\chem{O_2}$a${\text{(0-0)}} is partly compensated by a~simultaneous
rise of the \chem{OH} layer of 1.1\,\unit{km} between the first and
last period. The smallest discrepancy and scatter of $0.2$ and
$1.1$\,\unit{km} for \chem{O_2}$a${\text{(0-0)}} are found for the
fourth nighttime interval. The result for the discrepancy will change
with a~different \chem{OH}($v^{\prime}$). The analysis of the
$h_{\text{eff}}$ data confirms our expectation that an~\chem{OH} and
\chem{O_2}$a${\text{(0-0)}} $T_{\text{rot}}$ comparison is possible
with only minor corrections for a~wide range of nighttimes (see
Sect.~\ref{sec:tempcorrect}). A~correction is not negligible even for
equal $h_{\text{eff}}$ since the profile shapes and FWHM can be very
different (see Fig.~\ref{fig:rVER_all}). The average emission profile
FWHM for \chem{O_2}$a${\text{(0-0)}} and \chem{OH} are 12.9 and
8.7\,\unit{km}, respectively. There is still a~difference of
3.6\,\unit{km} (corresponding to 11.7 and 8.0\,\unit{km}) for the
fourth nighttime period, which shows the best agreement in terms of 
$h_{\text{eff}}$. The emission profile widths for 
\chem{O_2}$a${\text{(0-0)}} and \chem{OH} decrease during the night.
The difference is 4.6 and 2.3\,\unit{km}, respectively, if the FWHM of 
the first and last period are compared.

The emission profiles of \chem{O_2}$a${\text{(0-0)}} and the
negligible non-LTE contributions to the corresponding $T_{\text{rot}}$
make this band ideal for an~analysis of the \chem{OH}-related
$T_{\text{rot}}$. However, the strong profile variations do not allow
one to assign a~reference altitude for the
\chem{O_2}$a${\text{(0-0)}}-related temperatures. Therefore, we also
introduce a~reference profile, which is based on
\chem{O_2}$a${\text{(0-0)}} profiles observed in the second half of
the night. Since the corresponding median $h_{\text{peak}}$ and FWHM
are 89.8 and 10.9\,\unit{km}, respectively, we selected 90.0 and
11.0\,\unit{km} as reference values. Observed SABER VER profiles with
similar properties were then averaged. As the Cerro Paranal SABER
sample is too small for this, we used the larger sample without
longitude restriction (see Sect.~\ref{sec:saber}). We considered all
297 profiles with $h_{\text{peak}}$ between 88 and 90\,\unit{km} and
FWHM between 10.8 and 11.2\,\unit{km}. To avoid any broadening of the
averaged profile, the limits for the FWHM were more restrictive. In
the case of $h_{\text{peak}}$, this effect could be prevented by
shifting the profiles to the reference height before the
averaging. Figures~\ref{fig:rVER_all} and \ref{fig:rVERO2a_nbina} show
the resulting reference profile, which is not significantly affected
by the decaying daytime $a^1\Delta_{\mathrm{g}}$ population. The profile
is slightly asymmetric with $h_{\text{cen}} = 90.1$\,\unit{km} and
$h_{\text{eff}} = 90.4$\,\unit{km}.

\subsection{Temperature corrections}\label{sec:tempcorrect}

The final step of the preparation of $T_{\text{rot}}$ data for
a~comparison is the correction of differences in the emission
profiles. This requires some knowledge of the true temperature profile
in the mesopause region. SABER provides this kind of data
($T_{\text{kin}}$), which are based on \chem{CO_2}-related
measurements and non-LTE radiative transfer calculations (see
Sect.~\ref{sec:saber}). For the derivation of the sample-averaged
temperature differences for slightly deviating emission profiles, the
$T_{\text{kin}}$ uncertainties are probably small. However, we also
want to directly compare X-shooter $T_{\text{rot}}$ and SABER
$T_{\text{kin}}$. For this reason, the temperature uncertainties have
to be known in detail. According to \citet{REZ15} and
J.~M.~Russell~III (personal communication, 2015), the total
uncertainties for single profiles are about 5\,\unit{K} at 
90\,\unit{km} and tend to increase with altitude. Since we are 
interested in average properties, we can neglect the statistical 
errors, which leaves systematic uncertainties of 3 to 4\,\unit{K} in 
the relevant altitude range. This may further decrease if it is 
considered that systematic errors can also partly cancel out 
\citep{REZ15}.  For this reason, we assume a~relevant uncertainty of 
2.5\,\unit{K} as input for Sect.~\ref{sec:results}. As an additional 
check, we compared the preferred version 2.0 data to those of version 
1.07, which is possible for SABER data taken until the end of 2012. 
The old $T_{\text{kin}}$ show deviations from the new ones, which are
between $+2.0$\,\unit{K} for \chem{OH}($v^{\prime} = 2$) and 
$-0.2$\,\unit{K} for \chem{O_2}$b${\text{(0-1)}}. Assuming that the 
new $T_{\text{kin}}$ are more correct, the changes are consistent with 
the assumed errors.

To obtain effective temperatures $T_{\text{eff}}$ from the SABER
$T_{\text{kin}}$ for the different bands, we used a~similar procedure
as described for $h_{\text{eff}}$ (Sects.~\ref{sec:OHprof} and
\ref{sec:O2aprof}), i.e. the $T_{\text{kin}}$ of the different
altitudes were weighted depending on the related VERs for each
emission layer. Altitudes with negative VERs were rejected. Moreover,
weights were set to zero for altitude distances of more than
15\,\unit{km} from the \chem{OH} $h_{\text{cen}}$. 

The difference between the resulting $T_{\text{eff}}$ of a~given band 
and a~desired reference band $\Delta T_{\text{eff}}$ indicates the 
change of the effective true temperature by deviating altitudes, 
widths, and shapes of the corresponding emissions profiles. For the
derivation of possible non-LTE effects, these 
$\Delta T_{\text{eff}}$ have to be subtracted from the measured 
$T_{\text{rot}}$. The corrections cannot directly be applied since
both quantities originate from different data sets, i.e the Cerro 
Paranal SABER sample with 1685 profiles (Sect.~\ref{sec:saber}) and 
the X-shooter sample with 343 spectra (Sect.~\ref{sec:xshooter}). For 
this reason, we needed to obtain SABER-specific parameters such as
$T_{\text{eff}}$ and $h_{\text{eff}}$ for each X-shooter spectrum. We
achieved this by smoothing the SABER data in the two dimensions day of
year (DOY) and time. As a~smoothing filter, we used a~2-D Gaussian
with $\sigma$ of half a~month (15.2~days) and half an~hour. The
Gaussians were calculated across the New Year boundary if
required. For each X-shooter data point, we could then compute weights
for the different SABER data points and, finally, weighted averages
for the SABER-related parameters.

A~suitable smoothing procedure is crucial since both data sets have
large gaps. In particular, the SABER data for our selected area around 
Cerro Paranal only cover a~very narrow range of times for a~certain 
DOY (see Sect.~\ref{sec:saber}). It is desirable to average as many 
SABER data as possible to obtain reliable mean values. However, if 
this requires to add data taken at very different conditions, this 
might cancel out variability patterns that are required for realistic 
corrections. The SABER data set is already a~compromise since it 
includes data that were taken several 100\,\unit{km} away from Cerro 
Paranal. Hence, we checked the influence of the
smoothing procedure on the resulting weighted parameters. Focusing on
the X-shooter sample mean $T_{\text{eff}}$ for the eight \chem{OH},
two \chem{O_2}, and one reference emission layers (see
Sect.~\ref{sec:profiles}), we found a~mean change of about
0.1\,\unit{K} for a~bisection of the Gaussian $\sigma$. A~doubling of
the limiting periods resulted in a~corresponding mean shift of about
0.2\,\unit{K}. Moreover, a~different smoothing function
(e.g. an~exponential function with a~flat top for distances smaller
than $\sigma$) had little influence on the results. Finally, the
introduction of the solar activity (measured by the solar radio flux)
as a~third dimension did not significantly change the resulting
$T_{\text{eff}}$ either. Therefore, we conclude that the details of the
smoothing procedure for the SABER data are not critical for our
analysis.

The main contributions to the weighted parameter averages mainly
originate from a~relatively small fraction of the SABER data. Within
the $1\sigma$ perimeter of the 2-D Gaussian around the Cerro Paranal 
data points, the number of SABER profiles ranges from 0 to 41 for our 
applied procedure. The mean is 12. If a~weight sum is used (defining 
weight~$= 1$ in the centre of the Gaussian), values between 4 and 49 
with a~mean of 23 are found. These numbers suggest that 
profile-to-profile variations limit the accuracy of the resulting 
averages. This is confirmed by typical $T_{\text{eff}}$ mean errors 
between 1.3\,\unit{K} (\chem{O_2}$a${\text{(0-0)}}) and 1.9\,\unit{K}
(\chem{O_2}$b${\text{(0-1)}}) for individual X-shooter
observations. These uncertainties are comparable to the systematic
$T_{\text{rot}}$ measurement errors (see
Sect.~\ref{sec:lines}). However, for sample averages, the statistical
$T_{\text{eff}}$-related errors strongly decrease. The full sample
mean errors are only about 0.1\,\unit{K}. The uncertainties are
further reduced if the $\Delta T_{\text{eff}}$ for the
$T_{\text{rot}}$ correction are taken into account. In this case, the
individual mean errors range from 0.6\,\unit{K}
(\chem{O_2}$a${\text{(0-0)}}) to 1.3\,\unit{K}
(\chem{O_2}$b${\text{(0-1)}}) for the reference profile at
90\,\unit{km} described in Sect.~\ref{sec:O2aprof}. The uncertainties
for the sample mean are negligible.

The most critical source of error is probably a~significant deviation
of the $T_{\text{eff}}$ climatologies from those related to the Cerro
Paranal X-shooter data. This is relevant since the set of SABER
profiles used covers a~relatively wide area comprising 3{\degree} in
latitude and 20{\degree} in longitude (see
Sect.~\ref{sec:saber}). Moreover, there are large gaps in the local
time coverage for a~given month. In order to investigate this, we used
the five nighttime periods first introduced in
Fig.~\ref{fig:rVERO2a_nbina}. For the \chem{OH}($v^{\prime}$) and two
\chem{O_2} bands, we then measured the sample-averaged difference of 
the X-shooter-related $T_{\text{rot}}$ and the SABER-related 
$T_{\text{eff}}$ for each time interval. Finally, mean values and 
standard deviations were calculated for the five data points. This 
provides us with a~rough estimate of the agreement of the X-shooter 
and SABER nocturnal variability patterns. The resulting standard 
deviations range from 0.7\,\unit{K} for \chem{OH}($v^{\prime} = 9$) to 
2.1\,\unit{K} for \chem{OH}($v^{\prime} = 2$). The average for all 
emission layers is 1.5\,\unit{K}. We can repeat this calculation by 
taking the temperature differences for two bands, which is needed for 
the emission profile correction. For the emission of
\chem{O_2}$a${\text{(0-0)}} as reference profile, the standard 
deviations now range from 0.2\,K for \chem{OH}($v^{\prime} = 3$) to 
3.2\,K for \chem{O_2}$b${\text{(0-1)}}, which reflects the relatively
low effective height of the reference layer. Our preferred reference 
profile at 90\,\unit{km} cannot be used in this context since 
$T_{\text{rot}}$ measurements do not exist. For this reason, we 
calculated the standard deviations for all possible band combinations 
and performed a~regression analysis depending on the related 
height differences $\Delta h_{\text{eff}}$. Then, the 
climatology-related uncertainties can be estimated only based on 
differences of emission altitudes. This approach assumes that 
$h_{\text{eff}}$ is the major driver for the changes in the 
variability pattern, which is likely since the widths of the emission 
layers do not differ a~lot. Moreover, a~linear change of the 
variability with $\Delta h_{\text{eff}}$ is required, which is more 
difficult to fulfil if the height differences are relatively large. 
Significant variations of the \chem{OH} non-LTE effects could 
influence the fit (see Sect.~\ref{sec:results}). For our X-shooter 
sample, we obtain a~slope of 0.37\,\unit{K\,km^{-1}} and an~offset of 
0.47\,\unit{K}. The latter includes influences like FWHM differences. 
The fit is robust since the uncertainties of both values are lower by 
more than an~order of magnitude. For a~change from a~band-specific 
emission profile to our reference profile, the linear fit results in 
systematic errors in the temperature correction by the different 
climatologies for the SABER and X-shooter data between 0.7\,\unit{K} 
for \chem{OH}($v^{\prime} = 9$) and 2.0\,\unit{K} for 
\chem{O_2}$b${\text{(0-1)}}. For the discussion of seasonal variations 
in Sect.~\ref{sec:seasvar}, we also performed the whole analysis by 
using double month periods, which have a~similar 
length as the TIMED yaw cycle of about 60~days \citep{RUS99}. 
Independent of the month grouping starting with December/January or 
January/February, we obtained slightly higher errors between 
1.1\,\unit{K} for \chem{OH}($v^{\prime} = 9$) and 2.3\,\unit{K} for
\chem{O_2}$b${\text{(0-1)}}. Our estimates might be lower than the
real systematic uncertainties due to the low bin number, which is 
limited by the small data sets and gaps in the time coverage
(cf. Sect.~\ref{sec:seasvar}).

The relatively large distances between Cerro Paranal and the tangent
points where the SABER profiles were taken are an~important reason for
the deviating variability. With the approach described above, we can
check the robustness of our area selection limits. Focusing on the
$|\lambda - \lambda_{\text{CP}}| \le 10${\degree} criterion
(Sect.~\ref{sec:saber}), we also studied 5{\degree} and 20{\degree} 
limits. For the correction of the differences between the emission 
profiles of the measured bands and the reference profile, we find 
average temperature uncertainties of 1.2, 1.3, and 1.4\,\unit{K} for 
the 5{\degree}, 10{\degree}, and 20{\degree} limits, respectively. 
The averages were derived from the results for the eight \chem{OH} and 
two \chem{O_2} emission layers. The errors show the expected increase 
with growing longitude interval. However, the changes are small enough 
that the selection of the limit does not need to be very accurate. 
A~tight longitude range is not recommended since this would 
significantly shrink the SABER sample and, hence, increase the 
statistical errors.

\subsection{Emission heights and temperature correlations}\label{sec:correl}

\begin{figure}[t]
\includegraphics[width=80mm,clip=true]{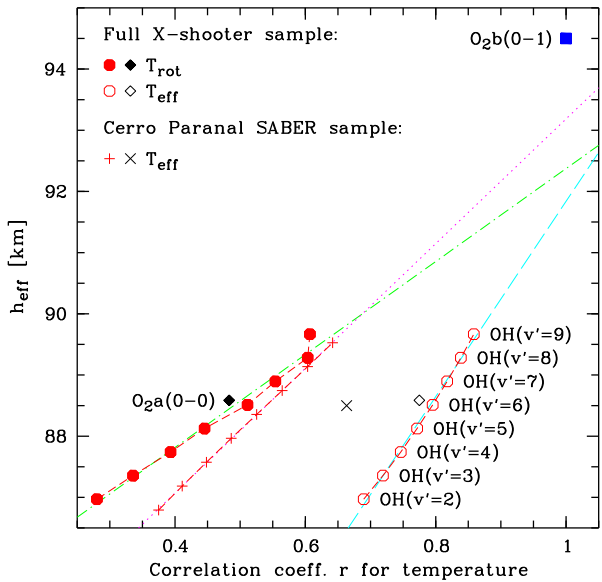}
\caption{Effective emission height $h_{\text{eff}}$ in \unit{km} from
  SABER data vs. the correlation coefficient $r$ for the correlation
  of the $T_{\text{rot}}$ (filled symbols), $T_{\text{eff}}$ for the
  full X-shooter sample (open symbols), and $T_{\text{eff}}$ for the
  Cerro Paranal SABER sample (crosses) of \chem{OH}($v^{\prime}$) and
  \chem{O_2}$a${\text{(0-0)}} with those of
  \chem{O_2}$b${\text{(0-1)}}. The dash-dotted ($T_{\text{rot}}$,
  X-shooter sample), long-dashed ($T_{\text{eff}}$, X-shooter sample),
  and dotted line ($T_{\text{eff}}$, SABER sample) show fits of the
  \chem{OH} data, which are connected by short-dashed lines.}
\label{fig:heff_r}
\end{figure}

In the same way as described in Sect.~\ref{sec:tempcorrect} for
$T_{\text{eff}}$, the band-specific $h_{\text{eff}}$
(Sect.~\ref{sec:OHprof}) were derived for each X-shooter
observation. It is possible to check the reliability of these data by
comparing average $h_{\text{eff}}$ with correlation coefficients $r$ 
for the correlation of the corresponding $T_{\text{rot}}$ with the  
$T_{\text{rot}}$ for a~reference band derived from the same data set
(Sect.~\ref{sec:lines}). Assuming a~gradual change of the temperature
variability pattern with height, the $T_{\text{rot}}$-based $r$ should
be a~function of the altitude difference of the emission layers used
for the correlation analysis. The required altitude-dependent
temperature variations can be caused by waves like tides
\citep[e.g.][]{MAR06}. A~complication could be $v^{\prime}$-dependent
variations of the non-LTE contributions to the measured \chem{OH}
$T_{\text{rot}}$ (see Sect.~\ref{sec:results}). As $r$ is sensitive to
measurement uncertainties, we only selected the most reliable
\chem{OH} band for each $v^{\prime}$, i.e. we used {\text{(2-0)}},
{\text{(3-1)}}, {\text{(4-2)}}, {\text{(5-3)}}, {\text{(6-2)}},
{\text{(7-4)}}, {\text{(8-5)}}, and {\text{(9-7)}}. The results for
the latter band are only based on 240 instead of 343 spectra due to
a~lack of wavelength coverage and issues with the flux calibration
(see Sect.~\ref{sec:xshooter}).

Figure~\ref{fig:heff_r} shows the mean $h_{\text{eff}}$ vs. $r$ for
the X-shooter sample with \chem{O_2}$b${\text{(0-1)}} as a~reference
band ($r = 1$). There is a~striking linear relation for the \chem{OH}
data, which supports our assumption of equidistant emission layers
(Sect.~\ref{sec:OHprof}). The slope of a~linear fit is
$7.6\,{\pm}\,0.4$\,\unit{km}. The deviation of \chem{OH}{\text{(9-7)}}
may be explained by the different sample and the measurement
uncertainties. \chem{O_2}$a${\text{(0-0)}} is very close to the fit
and best agrees with \chem{OH}{\text{(6-2)}}. This is another argument
for using this band for temperature comparisons with \chem{OH}
data. \chem{O_2}$b${\text{(0-1)}} deviates by about 2\,\unit{K} from
the extrapolated fit, which is most probably caused by a~non-linear
relation of $h_{\text{eff}}$ and $r$ for the large emission height
differences. The mesopause temperature minimum, which is close to the
\chem{O_2}$b${\text{(0-1)}} emission layer (see
Sect.~\ref{sec:results}), might have an~impact. Our interpretations
are supported by the $r$ that we derived from the SABER-related
$T_{\text{eff}}$ for the X-shooter sample
(Fig.~\ref{fig:heff_r}). There is an~almost perfect linear relation of
$h_{\text{eff}}$ and $r$ for these data, which is in good agreement 
with the found equidistance of the \chem{OH} layers and the position 
of \chem{O_2}$a${\text{(0-0)}} in relation to the \chem{OH} bands. The
position of \chem{O_2}$b${\text{(0-1)}} deviates in a~similar way from
a~linear fit of the \chem{OH} data with a~slope of $16.0 \pm
0.4$\,\unit{km}. The correlations for $T_{\text{eff}}$ are
significantly stronger than for $T_{\text{rot}}$. The relatively low
signal-to-noise ratio of the \chem{O_2}$b${\text{(0-1)}} data compared
to those of the other bands can have a~small impact. However, a~direct
correlation for e.g. $v^{\prime} = 3$ and 8 revealed $r = 0.86$ for
$T_{\text{rot}}$ and $0.97$ for $T_{\text{eff}}$. Possible variations
of the \chem{OH} $T_{\text{rot}}$ non-LTE contributions do not appear
to be sufficiently strong since the $r$ difference for
\chem{O_2}$a${\text{(0-0)}} is in good agreement with those of
\chem{OH} (cf. Sect.~\ref{sec:noctvar}). Therefore, the best
explanation for the $r$ discrepancies is the weighted averaging of the
SABER measurements to derive representative values for the X-shooter
data set (Sect.~\ref{sec:tempcorrect}). This effect can be estimated
by calculating $r$ for the $T_{\text{eff}}$ of the original Cerro
Paranal SABER sample with 1685 profiles (Sect.~\ref{sec:saber}). As
shown in Fig.~\ref{fig:heff_r}, the \chem{OH} $r$ values for these
data with a~slope of $10.2 \pm 0.1$\,\unit{km} are very close to those
for $T_{\text{rot}}$, i.e. the $r$ differences are indeed related to
the averaging procedure. The position of \chem{O_2}$a${\text{(0-0)}}
deviates from the trend for \chem{OH}, which is probably caused by
differences in the X-shooter and SABER sample properties. The latter
also contributes to the remaining differences between the \chem{OH}
data points. Moreover, the limited vertical resolution of the SABER
profiles can cause higher $r$ for $T_{\text{eff}}$ than for
$T_{\text{rot}}$. The sampling of about 0.4\,\unit{km}
(Sect.~\ref{sec:saber}) is similar to the height difference of
adjacent \chem{OH}($v^{\prime}$) emission layers
(Sect.~\ref{sec:OHprof}). Therefore, small-scale variations cannot be
probed by SABER.

\section{Results and discussion}\label{sec:results}

After the derivation of $T_{\text{rot}}$ corrected for the emission
layer differences (Sect.~\ref{sec:analysis}), we can finally compare
these temperatures and quantify the \chem{OH}-related non-LTE effects
depending on $v^{\prime}$ and time.

\subsection{Temperature comparison for reference profile}\label{sec:tempcomp}

\begin{figure}[t]
\includegraphics[width=80mm,clip=true]{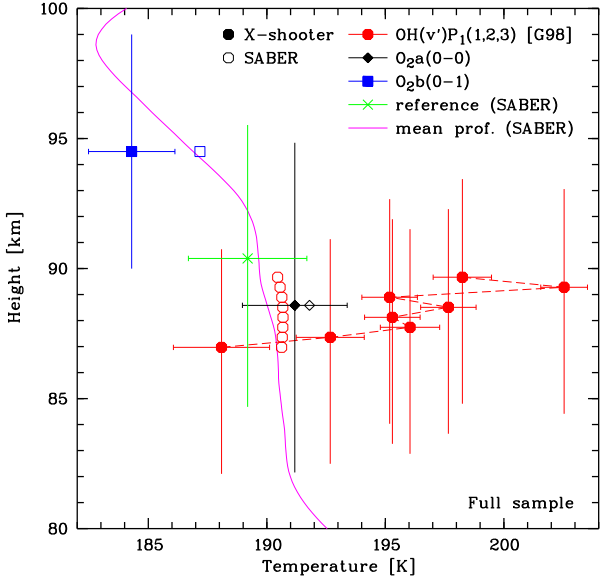}
\caption{Comparison of temperatures in \unit{K} and the corresponding
  heights in \unit{km} for the full X-shooter sample. Solid symbols
  show X-shooter $T_{\text{rot}}$ measurements and their uncertainties
  (excluding errors related to the molecular parameters where HITRAN
  was used) for \chem{OH} from $v^{\prime} = 2$ to 9
  (cf. Fig.~\ref{fig:heff_r}) using the first three $P_1$ lines
  (circles), \chem{O_2}$a${\text{(0-0)}} (diamond), and
  \chem{O_2}$b${\text{(0-1)}} (square). The effective emission heights
  were derived by means of SABER data. The bars in height direction
  mark the profile width at half maximum. The cross is related to the
  SABER $T_{\text{kin}}$ for the \chem{O_2}$a${\text{(0-0)}}-based
  reference profile. Open symbols indicate the $T_{\text{kin}}$
  results for the emission profiles of the bands measured by
  X-shooter. For these data, we assume temperature uncertainties of
  2.5\,\unit{K}. The solid curve shows the mean $T_{\text{kin}}$
  profile for the X-shooter sample.}
\label{fig:Trot_h}
\end{figure}

Figure~\ref{fig:Trot_h} gives an~overview of the mean $T_{\text{rot}}$
and their uncertainties for the eight \chem{OH}($v^{\prime}$),
\chem{O_2}$b${\text{(0-1)}}, and \chem{O_2}$a${\text{(0-0)}} derived
from the full X-shooter sample of 343 spectra (see
Sect.~\ref{sec:lines}). The $T_{\text{rot}}$ are plotted in comparison
to the corresponding SABER-based $h_{\text{eff}}$
(Sect.~\ref{sec:profiles}) projected onto the X-shooter sample
(Sect.~\ref{sec:tempcorrect}). The FWHM of the emission profiles are
indicated by vertical bars. The average \chem{O_2}$a${\text{(0-0)}}
profile is the widest with about 12.7\,\unit{km}. The figure shows
a~large overlap of the \chem{OH} and \chem{O_2}$a${\text{(0-0)}}
emission profiles, as already demonstrated by
Fig.~\ref{fig:rVER_all}. However, there are large differences in the
measured \chem{OH} $T_{\text{rot}}$, which range from 188.1\,\unit{K}
for $v^{\prime} = 2$ to 202.5\,\unit{K} for $v^{\prime} = 8$. As
discussed in \citet{NOL15} and Sect.~\ref{sec:intro}, the maximum at
$v^{\prime} = 8$ is typical of the $T_{\text{rot}}$ ($v^{\prime}$)
pattern and can be explained by the contribution of a~nearly nascent
population, which shows a~particularly hot rotational level population
distribution for this $v^{\prime}$. The mean $T_{\text{rot}}$ of
\chem{O_2}$a${\text{(0-0)}} is at the lower end of the \chem{OH}
$T_{\text{rot}}$ distribution with $191.2 \pm 2.2$\,\unit{K}. The
\chem{O_2}$b${\text{(0-1)}} $T_{\text{rot}}$ of $184.3 \pm
1.8$\,\unit{K} is even lower. These temperatures can be compared to
the SABER-based $T_{\text{eff}}$ for the X-shooter sample with
an~estimated uncertainty of about 2.5\,\unit{K}
(Sect.~\ref{sec:tempcorrect}). While the $T_{\text{rot}}$ and
$T_{\text{eff}}$ for each \chem{O_2} band agree within the errors,
this is not the case for all the \chem{OH} bands. The $v^{\prime}$
dependence of the temperature is also very different. The \chem{OH}
$T_{\text{eff}}$ only range from 190.5 to 190.7\,\unit{K}, i.e. the
temperature gradient is almost zero. In Fig.~\ref{fig:Trot_h}, we also
display the mean $T_{\text{kin}}$ profile calculated for the X-shooter
sample. Indeed, the mean temperature profile is relatively flat in the
altitude range of the \chem{OH} emission. Together with the large
profile overlap, this explains the almost absent temperature
gradient. This implies that the contribution of real temperature
variations to the characteristic $T_{\text{rot}}$ ($v^{\prime}$)
pattern is small. Non-LTE effects have to dominate. Only the
relatively low $T_{\text{rot}}$ of \chem{O_2}$b${\text{(0-1)}} can be
explained by the proximity of the mesopause temperature minimum.

\begin{figure}[t]
\includegraphics[width=80mm,clip=true]{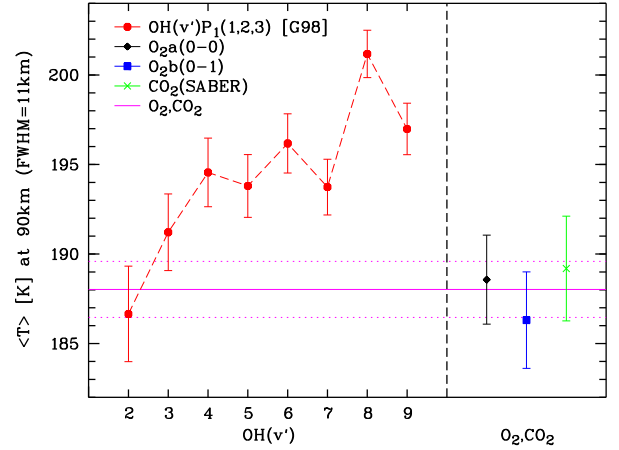}
\caption{Comparison of mean temperatures in \unit{K} for the
  \chem{O_2}$a${\text{(0-0)}}-based reference profile and the full
  X-shooter sample. The $v^{\prime}$-dependent \chem{OH}
  $T_{\text{rot}}$ (circles) are based on measurements of the first
  three $P_1$ lines and HITRAN-based molecular parameters
  \citep{GOL98}. The systematic errors related to the latter are not
  included in the plotted error bars. HITRAN-based molecular
  parameters were also used for the derivation of the $T_{\text{rot}}$
  from \chem{O_2}$a${\text{(0-0)}} (diamond) and
  \chem{O_2}$b${\text{(0-1)}} (square). The cross shows the
  \chem{CO_2}-based SABER $T_{\text{kin}}$ for the reference profile
  and the full X-shooter sample. The mean and the uncertainties
  derived from the three non-\chem{OH} temperatures are marked by
  horizontal lines.}
\label{fig:Trotref_hit}
\end{figure}

Figure~\ref{fig:Trot_h} also shows the $h_{\text{eff}}$, FWHM, and
$T_{\text{eff}}$ of the \chem{O_2}$a${\text{(0-0)}}-based reference
profile introduced in Sect.~\ref{sec:O2aprof}. $T_{\text{eff}}$ is
$189.2 \pm 2.5$\,\unit{K}. The different $T_{\text{rot}}$ can be 
corrected to be representative of this profile that peaks at 
90\,\unit{km} and has a~FWHM of 11\,\unit{km}. This can be done by
subtracting the difference between the $T_{\text{eff}}$ related to a
given band and the reference profile from the $T_{\text{rot}}$ for the
selected band (see Sect.~\ref{sec:tempcorrect}). The resulting 
temperatures are shown in Fig.~\ref{fig:Trotref_hit}. For 
\chem{O_2}$a${\text{(0-0)}} and \chem{O_2}$b${\text{(0-1)}}, we 
obtained $188.6 \pm 2.5$\,\unit{K} and $186.3 \pm 2.7$\,\unit{K}, 
respectively. The uncertainties are a~combination of the measurement 
uncertainties already shown in Fig.~\ref{fig:Trot_h} and the 
temperature correction errors discussed in 
Sect.~\ref{sec:tempcorrect}. The two \chem{O_2}-related temperatures 
and the \chem{CO_2}-based SABER $T_{\text{eff}}$ agree very well 
within their errors. Since these three temperatures are not affected 
by non-LTE effects, we can combine them, which results in an~average 
temperature of $188.0 \pm 1.6$\,\unit{K}. \citet{CLE11} published 
a~mean temperature profile based on sodium lidar measurements carried 
out at S\~ao Jos\'e dos Campos in Brazil (23{\degree}\,S, 
46{\degree}\,W) from 2007 to 2010. For our reference emission and 
their temperature profile, this implies a~$T_{\text{eff}}$ of 
188\,\unit{K}, which is in remarkable agreement with our measurement 
despite their assumed absolute errors of about 5\,\unit{K}, 
differences in the time coverage (no January data), and the small but 
probably non-negligible differences in latitude and longitude. For the 
latter effect, see Fig.~\ref{fig:Th_lon}. The upper part of the 
lidar-based temperature profile is also in good agreement with the 
SABER one in Fig.~\ref{fig:Trot_h}. The lower part shows some
discrepancies due to a~different height and strength of the secondary
minimum, which is at about 88\,\unit{km}. In the SABER case, there is
only a~nearly constant temperature down to about 81\,\unit{km}, which
causes a~lower onset of the steep temperature increase that is already
visible in the \citet{CLE11} profile above 80\,\unit{km}. The
comparison with satellite and lidar data shows that the
$T_{\text{rot}}$ from \chem{O_2}$a${\text{(0-0)}} and
\chem{O_2}$b${\text{(0-1)}} line measurements in X-shooter spectra
provide reliable temperatures in the mesopause region.

\begin{figure}[t]
\includegraphics[width=80mm,clip=true]{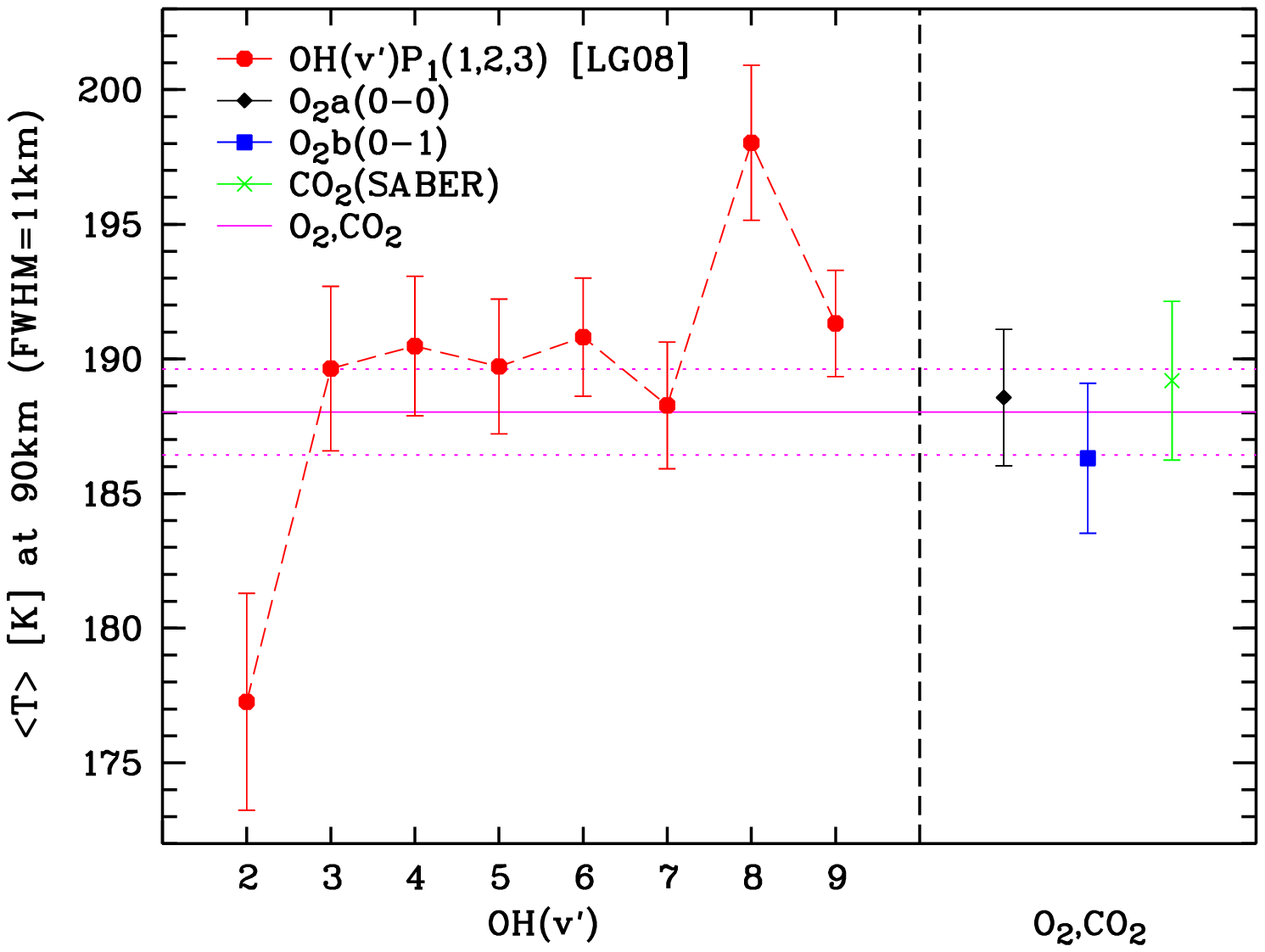}
\caption{Same plot as Fig.~\ref{fig:Trotref_hit} except for \chem{OH}
  $T_{\text{rot}}$ based on the molecular parameters of
  \citet{LOO07,LOO08}.}
\label{fig:Trotref_vdL}
\end{figure}

The \chem{OH} $T_{\text{rot}}$ indicate significant non-LTE
contributions in comparison to our LTE reference temperature of about
188\,\unit{K}. The temperatures range from $186.7 \pm 2.7$\,\unit{K}
for $v^{\prime} = 2$ to $201.2 \pm 1.3$\,\unit{K} for $v^{\prime} =
8$. In the latter case, the difference is about 13\,\unit{K}. With the
estimated errors, the $\Delta T_{\text{non-LTE}}$ of $v^{\prime} = 8$,
9, and 6 are highly significant ($> 3\sigma$). The situation for
$v^{\prime} = 2$ and 3 is unclear. The \chem{OH} bands {\text{(2-0)}}
and {\text{(3-0)}}, which by far show the lowest $T_{\text{rot}}$
\citep{NOL15}, are also affected by relatively large systematic
uncertainties. \chem{OH}{\text{(2-0)}} is partly absorbed by water
vapour (more than any other band considered) and
\chem{OH}{\text{(3-0)}} is partly affected by line blending and close
to the upper wavelength limit of the X-shooter VIS arm. Therefore, it
cannot be excluded that the true temperatures of $v^{\prime} = 2$ and
3 are similar to those of higher $v^{\prime}$. The plotted
temperatures are only valid if the $T_{\text{rot}}$ are derived from
the first three $P_1$-branch lines. For example, the resulting $\Delta
T_{\text{non-LTE}}$ and the significance of the non-LTE effects would
increase if $P_1$ (4) was also used. In this case, the
$T_{\text{rot}}$ would be higher by about 1\,\unit{K} for all
$v^{\prime}$. The change to the first four $P_2$-branch lines would
even increase the \chem{OH} $T_{\text{rot}}$ by about
11\,\unit{K}. Moreover, the discussed \chem{OH} $T_{\text{rot}}$ are
based on Einstein coefficients and level energies from the HITRAN
database \citep{ROT13}, which are related to \citet{GOL98}. The
results change if we use the molecular parameters of
\citet{LOO07,LOO08} (Fig.~\ref{fig:Trotref_vdL}). The $T_{\text{rot}}$
are more uncertain and tend to be lower than those based on HITRAN
data. If we neglect the very uncertain data point for $v^{\prime} = 2$
\citep[see][]{NOL15}, the mean shift is 4.2\,\unit{K}. Hence, only the
$\Delta T_{\text{non-LTE}}$ of 10\,\unit{K} for $v^{\prime} = 8$
remains significant with more than $3\sigma$. Although the
$T_{\text{rot}}$ based on \citet{LOO08} are less reliable than those
based on \citet{GOL98}, these results demonstrate that the
uncertainties of the molecular parameters limit the accuracy of
$\Delta T_{\text{non-LTE}}$ that can currently be
achieved. Nevertheless, $\Delta T_{\text{non-LTE}}$ of more than 
10\,\unit{K} for $v^{\prime} = 8$ are plausible if the first three 
$P_1$-branch lines are used. Finally, the results show that the 
consideration of emission profile differences does not significantly 
change the characteristic $T_{\text{rot}}$ ($v^{\prime}$) pattern 
found by \citet{COS07} and \citet{NOL15}. It can only be explained by 
non-LTE effects.

\subsection{Nocturnal variations of temperature differences}\label{sec:noctvar}

\begin{figure}[t]
\includegraphics[width=80mm,clip=true]{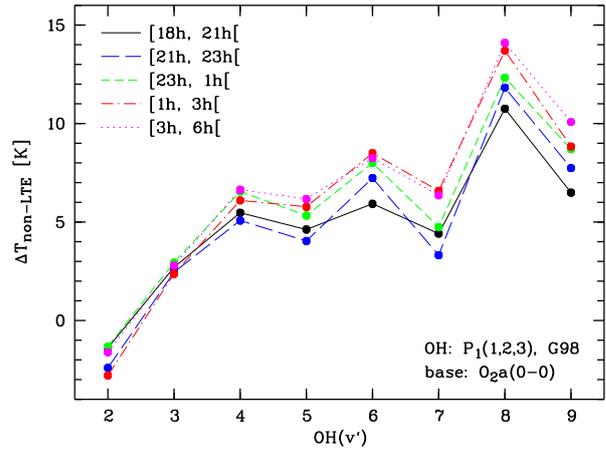}
\caption{$v^{\prime}$-dependent non-LTE contributions to \chem{OH}
  $T_{\text{rot}}$ based on the first three $P_1$ lines and HITRAN
  data \citep{GOL98} for five nighttime LT periods derived from the
  full X-shooter sample. The LTE base temperature for each
  $v^{\prime}$ was derived from the corresponding
  \chem{O_2}$a${\text{(0-0)}} $T_{\text{rot}}$ corrected for the
  difference in the VER profiles. The uncertainties of the absolute
  $\Delta T_{\text{non-LTE}}$ are similar to those given in
  Fig.~\ref{fig:Trotref_hit} (excluding the unknown uncertainties of
  the molecular parameters). Comparisons of $\Delta
  T_{\text{non-LTE}}$ for the different night-time periods and/or
  $v^{\prime}$ are more reliable. The errors should be of the order of
  1\,\unit{K}, i.e. very likely smaller than 2\,\unit{K}.}
\label{fig:delTrot_vp_nbina}
\end{figure}

\begin{figure}[t]
\includegraphics[width=80mm,clip=true]{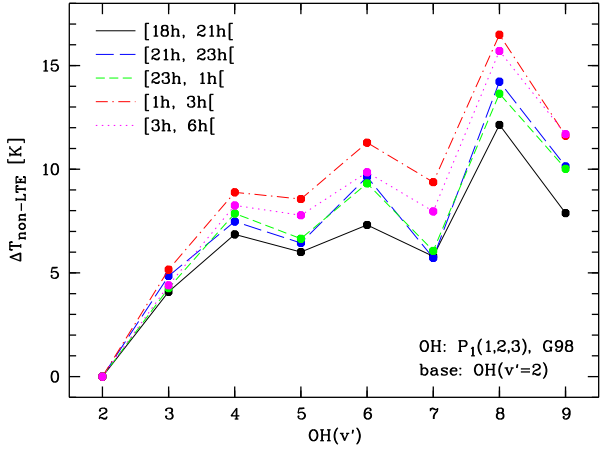}
\caption{Same plot as Fig.~\ref{fig:delTrot_vp_nbina} except for
  \chem{OH}($v^{\prime} = 2$) as base for the calculation of $\Delta
  T_{\text{non-LTE}}$. The errors for comparisons of different
  nighttime periods and/or $v^{\prime}$ are probably smaller than
  those of Fig.~\ref{fig:delTrot_vp_nbina} since temperature
  correction uncertainties are less critical. The absolute $\Delta
  T_{\text{non-LTE}}$ are less reliable.}
\label{fig:delTrot_vp_nbina_bv2}
\end{figure}

\citet{NOL15} studied the nocturnal variations of the \chem{OH}
$T_{\text{rot}}$ by means of five nighttime periods, which we also
used in Figs.~\ref{fig:rVERO2a_nbina} and~\ref{fig:heff_nbina}. The
analysis resulted in HITRAN-based $T_{\text{rot}}$ differences
(i.e. maximum minus minimum) between 4.8\,\unit{K} for $v^{\prime} =
5$ and 7.0\,\unit{K} for $v^{\prime} = 6$. In order to separate
non-LTE from temperature profile variations, we used the observed
\chem{O_2}$a${\text{(0-0)}} profiles as reference for an~emission
profile correction. We did not take the reference profile at
90\,\unit{km}, which was the basis for the discussion in
Sect.~\ref{sec:tempcomp}, since this reduces the uncertainties related
to the temperature correction from about 1.2 to 0.9\,\unit{K} on
average. At the beginning/end of the night, the uncertainties are
higher for high/low $v^{\prime}$. The resulting $\Delta
T_{\text{non-LTE}}$ are displayed in
Fig.~\ref{fig:delTrot_vp_nbina}. Compared to the uncorrected
$T_{\text{rot}}$, the time-related variation for fixed $v^{\prime}$
significantly decreased. The $\Delta T_{\text{non-LTE}}$ differences
range from 0.6\,\unit{K} for $v^{\prime} = 3$ to 3.6\,\unit{K} for
$v^{\prime} = 9$. Hence, the average nocturnal variations of the
non-LTE contributions to the \chem{OH} $T_{\text{rot}}$ are
significantly lower than their actual values. Moreover, at least the
low $v^{\prime}$ indicate that the $T_{\text{rot}}$ variation is
obviously dominated by the variability of the real temperature
profile. However, it seems that only one third of the $T_{\text{rot}}$
($v^{\prime} = 9$) variability can be explained in this way.
Figure~\ref{fig:delTrot_vp_nbina} indicates a~nocturnal trend of
increasing $\Delta T_{\text{non-LTE}}$. The trend appears to be
stronger for higher $v^{\prime}$. The significance is about
$3\sigma$ for the differences between the first and last interval
and is limited by the statistical mean errors with an average of about
1.1\,\unit{K} and at least part of the temperature correction errors
due to the changing \chem{O_2}$a${\text{(0-0)}} emission profile.

\citet{NOL15} discovered a~significant increase of the
$T_{\text{rot}}$ differences of adjacent intermediate $v^{\prime}$ for
the second compared to the first nighttime period. In
Fig.~\ref{fig:delTrot_vp_nbina}, this causes a~relatively strong
$\Delta T_{\text{non-LTE}}$ increase for $v^{\prime} = 6$ but
a~decrease for $v^{\prime} = 5$ and 7. In order to better evaluate the
reliability of these trends, we also calculated $\Delta
T_{\text{non-LTE}}$ based on \chem{OH}($v^{\prime} = 2$) instead of
\chem{O_2}$a${\text{(0-0)}} emission profiles. Although the absolute 
$\Delta T_{\text{non-LTE}}$ are less trustworthy and possible weak 
non-LTE $T_{\text{rot}}$ variations for $v^{\prime} = 2$ are 
neglected, this approach is justified for the comparisons of nighttime 
intervals and $v^{\prime}$ since it can further reduce the 
uncertainties related to the temperature correction. The emission 
profile differences between \chem{OH}($v^{\prime} = 2$) and the other 
\chem{OH} bands are much smaller and less variable than those related 
to \chem{O_2}$a${\text{(0-0)}}. The resulting
Fig.~\ref{fig:delTrot_vp_nbina_bv2} does not show a~$\Delta
T_{\text{non-LTE}}$ decrease between the first and second nighttime
period. However, significant increases appear to be limited to
$v^{\prime} = 6$, 8, and 9. For $v^{\prime} = 5$ and 7, the major
change happens between the third and fourth period, i.e. after
midnight. The last period indicates a~decrease of $\Delta
T_{\text{non-LTE}}$ for most $v^{\prime}$. For $v^{\prime} \ge 6$, the
$\Delta T_{\text{non-LTE}}$ increases between the first and fourth
period vary between 3.6\,\unit{K} ($v^{\prime} = 7$) and 4.3\,\unit{K}
($v^{\prime} = 8$). The significance is higher than for the results
based on \chem{O_2}$a${\text{(0-0)}} as reference band.

The found $\Delta T_{\text{non-LTE}}$ variations appear to be in
agreement with those of the \chem{OH} emission layer height (see
Sect.~\ref{sec:intro}). Up to the fourth nighttime period, the
sample-averaged $h_{\text{eff}}$ increases by 0.9\,\unit{km}
($v^{\prime} = 9$) to 1.7\,\unit{km} ($v^{\prime} = 2$). Afterwards
the emission layers sink again except for $v^{\prime} = 9$. See
Fig.~\ref{fig:heff_nbina} for the $h_{\text{eff}}$ trend related to
$v^{\prime} = 8$. As already discussed by \citet{NOL15} based on
$T_{\text{rot}}$ differences and vibrational level populations,
a~higher emission layer is in a~lower density environment, which
decreases the frequency of thermalising collisions by \chem{O_2}
\citep[e.g.][]{ADL97,SAV12,XU12} and, consequently, can increase the
$T_{\text{rot}}$ non-LTE effects. The SABER data show that the
$h_{\text{eff}}$ increase is more related to a~VER decrease at low
altitudes than an~increase in the upper part of the emission
layer. This trend mainly depends on changes in the distribution of
atomic oxygen (which is required for the \chem{O_3} production) by
chemical reactions and atmospheric dynamics
\citep{LOW96,MAR06,NIK07}. In order to better estimate the effect of
the VER profile change on the $T_{\text{rot}}$ non-LTE effects, we
calculated the fraction of \chem{OH} emission above a~certain altitude
as a~function of the nighttime period. For this, we used the original
VER profiles of the two SABER channels at 1.64 and 2.06\,\unit{\mu m}
corresponding to $v^{\prime} = 4.6$ and 8.3, respectively
(Sect.~\ref{sec:OHprof}). For a~minimum altitude of 95\,\unit{km}, we
found an~increase from 4.7\,{{\%}} (first period) to 7.3\,{{\%}}
(fourth period) for $v^{\prime} = 4.6$ and from 9.4\,{{\%}} to
12.0\,{{\%}} for $v^{\prime} = 8.3$. The differences are similar for
both SABER channels but the absolute values are higher for $v^{\prime}
= 8.3$. The latter effect and the relative change of the emission
fractions with time enhance with increasing minimum altitude. These
results show the large impact of a~relatively small rise of the
emission layer on the highest parts of the VER profile. To fully
understand Fig.~\ref{fig:delTrot_vp_nbina_bv2}, it will be necessary
to model \chem{OH} $T_{\text{rot}}$ depending on $v^{\prime}$ and
altitude, which has not been done so far. While there is some
knowledge on the nascent population distribution over the rotational
levels of the different $v$ \citep{LLE78}, it is not clear how this 
distribution is modified by collisional quenching.

\begin{figure}[t]
\includegraphics[width=80mm,clip=true]{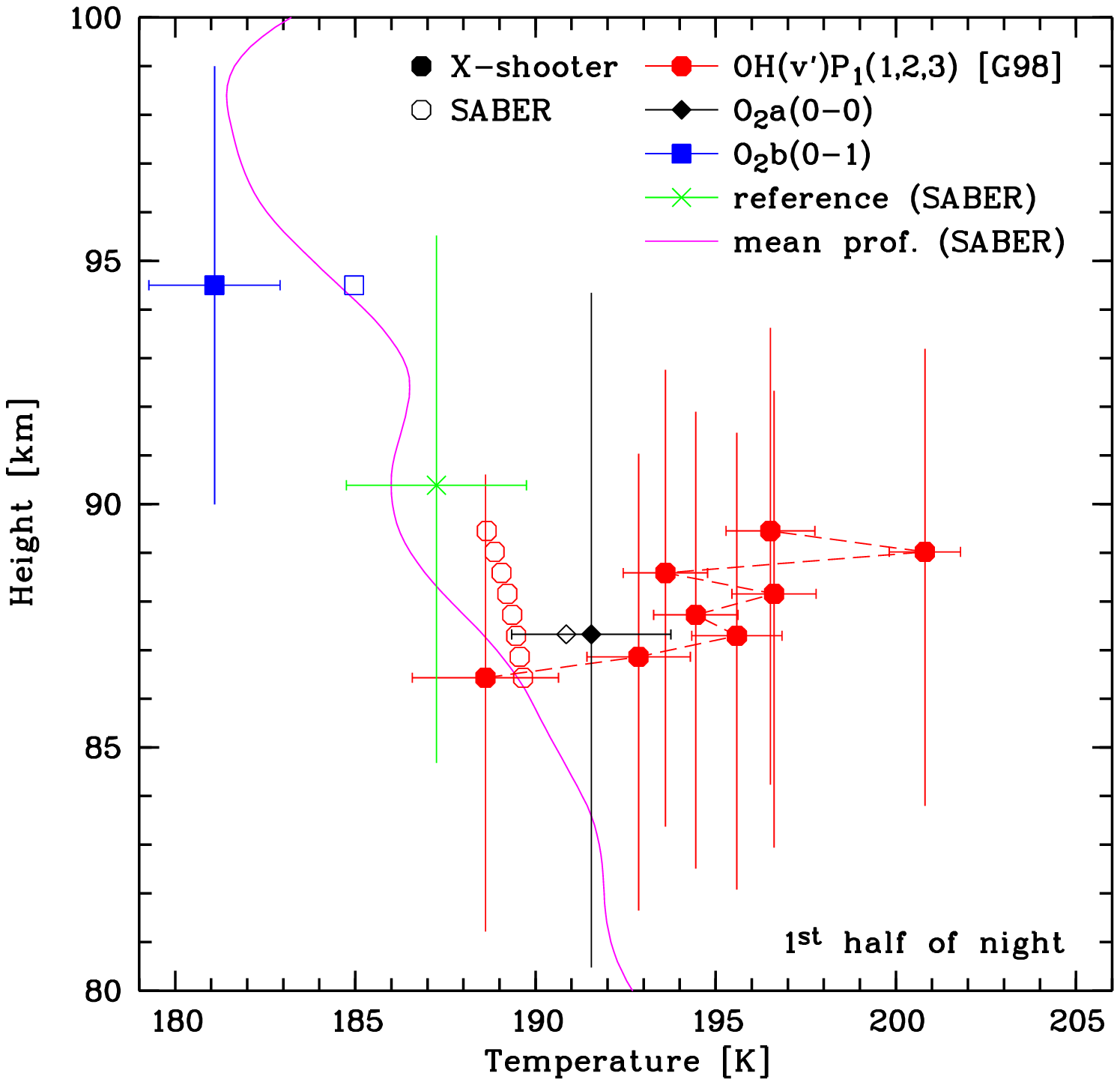}
\includegraphics[width=80mm,clip=true]{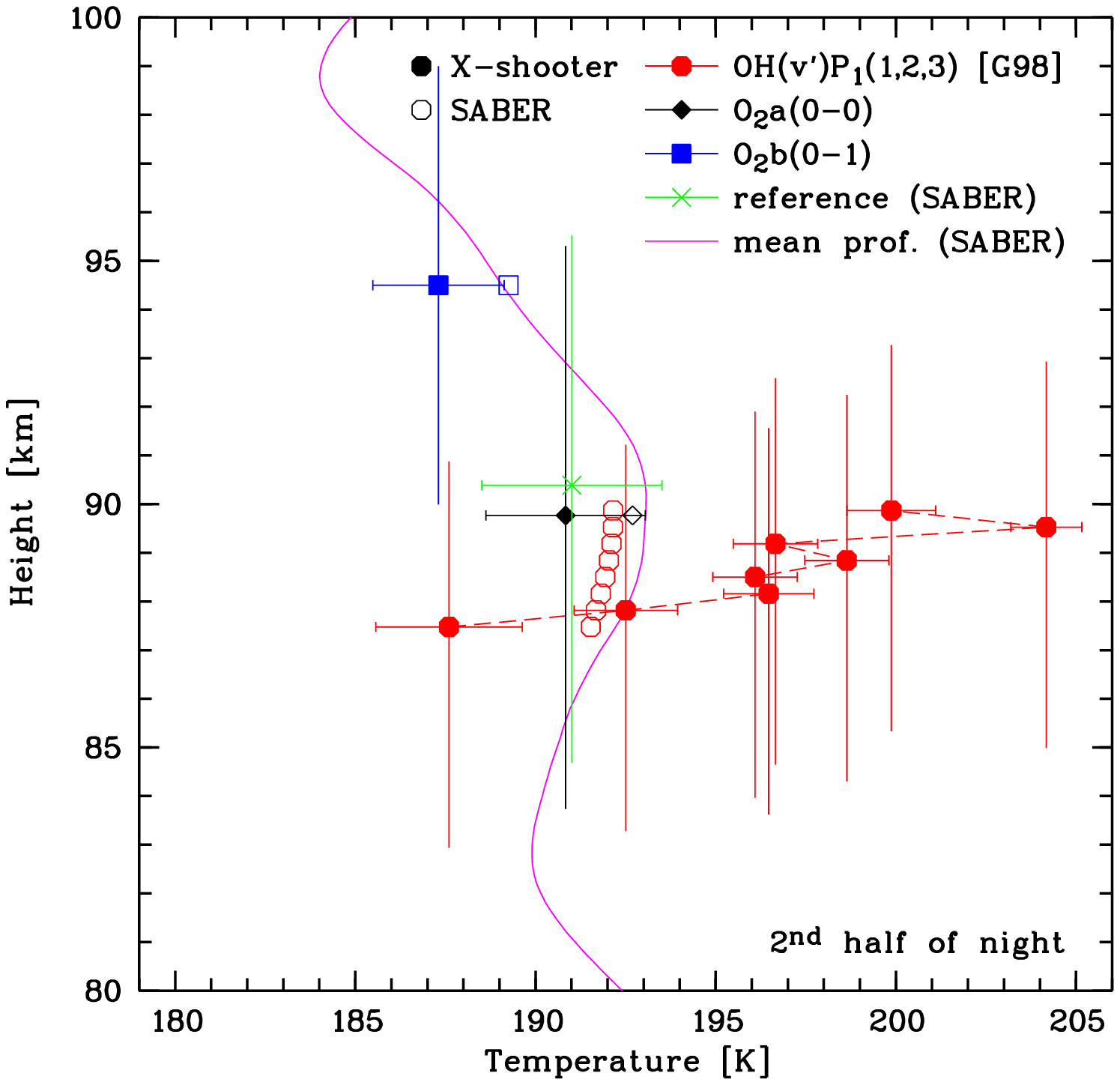}
\caption{Same plot as Fig.~\ref{fig:Trot_h} except for showing results
  for the first (upper panel) and the second half of the night (lower
  panel) separately.}
\label{fig:Trot_h_nh}
\end{figure}

Figure~\ref{fig:Trot_h_nh} illustrates the effect of the changing
emission layers and temperature profile on the measured
$T_{\text{rot}}$ by comparing $T_{\text{rot}}$, $T_{\text{eff}}$, and
$T_{\text{kin}}$ for the first and second half of the night. The plots
are similar to Fig.~\ref{fig:Trot_h} discussed in
Sect.~\ref{sec:tempcomp}. The \chem{OH} $T_{\text{rot}}$ show a~wider
spread in the second half of the night. This can partly be explained
by a~significant change of the temperature profile. The SABER data
reveal an average nocturnal trend from a~single deep mesopause
temperature minimum at the beginning to a~double minimum, where the
secondary trough is almost as deep as the primary one. The average
altitudes of both minima are 98 to 99\,\unit{km} and between 80 and
85\,\unit{km}. Observations of a~secondary minimum are not unusual
\citep{ZAH96,SHE04,FRI07,CLE11} and can be explained by tidal
perturbations and tide--gravity wave interactions
\citep{MERI00,MERI04}. The appearance of a~secondary mesopause minimum
causes a~temperature inversion at the main \chem{OH} emission
altitudes. The comparison of the second with the first half of the
night, therefore, shows a~$T_{\text{kin}}$ increase above and decrease
below 85\,\unit{km}. The \chem{OH} $T_{\text{eff}}$ are higher in the
second period and this increase is stronger for higher $v^{\prime}$
due to the different temperature gradients. The amounts are
1.9\,\unit{K} for $v^{\prime} = 2$ and 3.5\,\unit{K} for $v^{\prime} =
9$. Nevertheless, a~$T_{\text{rot}}$ increase with $v^{\prime}$
remains, which is disturbed by the odd/even $v^{\prime}$ dichotomy, as
indicated by Figs.~\ref{fig:delTrot_vp_nbina} and
\ref{fig:delTrot_vp_nbina_bv2}. The \chem{O_2}-related
$T_{\text{rot}}$ follow the changes of the temperature profile
convincingly. As expected, the emission height, profile width, and
temperature for \chem{O_2}$a${\text{(0-0)}} are much more similar to
those for the reference profile (Sect.~\ref{sec:O2aprof}) in the
second half of the night.

\subsection{Seasonal variations of temperature differences}\label{sec:seasvar}

\begin{figure}[t]
\includegraphics[width=80mm,clip=true]{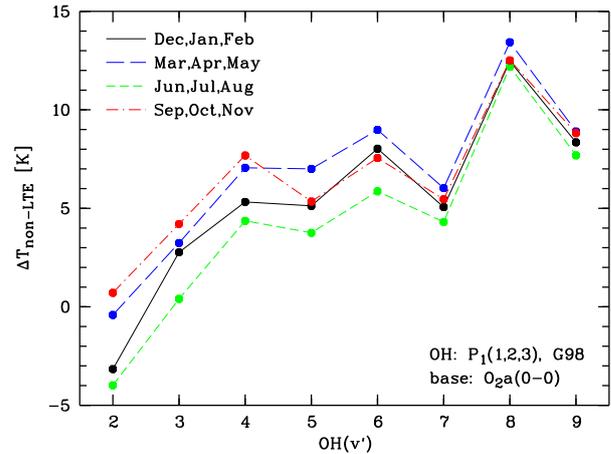}
\caption{Same plot as Fig.~\ref{fig:delTrot_vp_nbina} except for
  curves related to the four meteorological seasons instead of
  nighttime periods. The uncertainties for comparisons of $\Delta
  T_{\text{non-LTE}}$ for the different seasons and/or $v^{\prime}$
  are somewhat higher. 2\,\unit{K} and more cannot be excluded.}
\label{fig:delTrot_vp_seasa}
\end{figure}

\citet{NOL15} also studied the $T_{\text{rot}}$ ($v^{\prime}$) of the
four meteorological seasons. The X-shooter data indicated higher
temperatures for the equinoxes than for the solstices irrespective of
$v^{\prime}$. The maximum differences were between 7.4\,\unit{K} for
$v^{\prime} = 2$ and 3.3\,\unit{K} for $v^{\prime} = 8$. After
applying a~temperature correction based on
\chem{O_2}$a${\text{(0-0)}}, these differences decreased to 4.7 and
1.2\,\unit{K}, as shown in Fig.~\ref{fig:delTrot_vp_seasa}. These
values imply stronger non-LTE effects near the equinoxes than near the
solstices. Moreover, the effects seem to be stronger for lower
$v^{\prime}$. This is contradictory to the results and explanations
discussed in Sect.~\ref{sec:noctvar}. The apparent trend is most
probably caused by an~inefficient temperature correction. Although the
reduction of the seasonal variations by the conversion from
$T_{\text{rot}}$ to $\Delta T_{\text{non-LTE}}$ was even the largest
for low $v^{\prime}$ ($2.7$ vs. $2.1$\,\unit{K}, see above), it was
obviously insufficient to obtain reliable seasonal variations of the
\chem{OH} $T_{\text{rot}}$ non-LTE contributions. This is probably
related to the TIMED yaw cycle of 60~days, which causes large gaps in
the seasonal coverage for fixed local times (Sects.~\ref{sec:intro}
and \ref{sec:saber}). Therefore, the true seasonal variations could be
stronger than those used for the temperature correction. At Cerro
Paranal and similar latitudes, studying seasonal variations is more
challenging than an~analysis of the nocturnal variations since the
combination of semi-annual and the weaker annual oscillations for
\chem{OH} $T_{\text{rot}}$ \citep{TAK95,GEL08} is more complex than
the night trends \citep{TAK98,GEL08,NOL15}, which are dominated by the
solar migrating diurnal tide \citep{MAR06,ASMI12}. Hence, the
limitations of our SABER and X-shooter data sets do not allow us to
conclude on seasonal variations of the \chem{OH} $T_{\text{rot}}$
non-LTE effects.

\conclusions
\label{sec:conclusions}

Based on the VLT/X-shooter data set of \citet{NOL15} with 343 spectra,
we could compare their measurements of rotational temperatures
$T_{\text{rot}}$ from 25 \chem{OH} bands covering $v^{\prime} = 2$ to
9 with new results for the bands \chem{O_2}$b${\text{(0-1)}} and
\chem{O_2}$a${\text{(0-0)}}. For the latter band, it was just the
second time that $T_{\text{rot}}$ could be derived and the first time
that it could be done with spectroscopically resolved lines. Since
\chem{O_2}$a${\text{(0-0)}} emits at similar altitudes as \chem{OH}
and the rotational level populations appear to be consistent with
a~Boltzmann distribution, the accurate $T_{\text{rot}}$ for
\chem{O_2}$a${\text{(0-0)}} could be used to quantify the non-LTE
contributions to the \chem{OH} $T_{\text{rot}}$. To correct for the
different emission profiles and the corresponding VER-weighted
$T_{\text{eff}}$, we used \chem{OH}, \chem{O_2}$a${\text{(0-0)}}, and
\chem{CO_2}-based $T_{\text{kin}}$ profile data from TIMED/SABER
suitable for Cerro Paranal. The correction is particularly important
for the beginning of the night when the \chem{O_2}$a${\text{(0-0)}}
emission peaks several \unit{km} below the nighttime average due to
a~decaying daytime population from ozone photolysis. From the
X-shooter data set, we obtained an effective population lifetime of
about 46\,\unit{min}, which is in good agreement with SABER and other
data. The required emission profiles for the different \chem{OH}
$v^{\prime}$ were derived from the two \chem{OH} SABER channels under
the assumption that emission peaks are equidistant for adjacent
$v^{\prime}$. This approach is supported by a~comparison of the
effective emission heights $h_{\text{eff}}$ with temperature
correlation coefficients for the X-shooter and SABER samples, which
showed nearly linear relations.

For an~\chem{O_2}$a${\text{(0-0)}}-based reference profile at
90\,\unit{km}, we compared the corrected temperature measurements and
found that the results for \chem{O_2}$b${\text{(0-1)}} (where a~fixed
emission profile at 94.5\,\unit{km} was assumed),
\chem{O_2}$a${\text{(0-0)}}, and the SABER $T_{\text{kin}}$ agree very
well within the errors. The sample mean temperature at the reference
altitude was $188.0 \pm 1.6$\,\unit{K}. Except for the uncertain
$v^{\prime} = 2$ and 3, the resulting \chem{OH}-related temperatures
were significantly higher. For the first three $P_1$-branch lines and
HITRAN molecular data, a~maximum difference of about 13\,\unit{K} was
achieved for $v^{\prime} = 8$. The characteristic $T_{\text{rot}}$
($v^{\prime}$) pattern found by \citet{COS07} and \citet{NOL15}
showing an~odd/even $v^{\prime}$ dichotomy could also be seen in the
corrected temperature data. The amount of non-LTE contributions to the
\chem{OH} $T_{\text{rot}}$ $\Delta T_{\text{non-LTE}}$ strongly
depends on the selected lines. A~change of the line set affects all
\chem{OH} bands in a~similar way. Apart from measurement and
correction errors, the still uncertain molecular parameters limit the
accuracy of the derived $\Delta T_{\text{non-LTE}}$ to a~few \unit{K}.

An~analysis of the nocturnal variations of $\Delta T_{\text{non-LTE}}$
with five nighttime periods revealed that a~significant fraction of
the \chem{OH} $T_{\text{rot}}$ variability is caused by changes in the
emission and mesopause kinetic temperature profiles. However, 
fractions above 50\,{{\%}} could only be found for low $v^{\prime}$. 
At Cerro Paranal, the residual non-LTE effects tend to increase during 
the night and to be more variable at higher $v^{\prime}$. The largest 
$\Delta T_{\text{non-LTE}}$ differences were found for the first 
(before 21:00\,LT) and fourth period (between 01:00 and 03:00\,LT), 
and $v^{\prime} \ge 6$. They amount to about 4\,\unit{K} with an
uncertainty of about 2\,\unit{K}. The amplitude of the variations
appears to be only a~minor fraction of the total amount of the non-LTE
effects. The observed trends could be explained by the nocturnal rise
of the \chem{OH} emission layer, which we see in the SABER data up to
the fourth nighttime period, and a~corresponding reduction of the
thermalising collisions due to the lower effective \chem{O_2}
density. Higher $v^{\prime}$ could be more affected since the related
emission peaks at higher altitudes. We also studied possible seasonal
variations of $\Delta T_{\text{non-LTE}}$. However, the limitations of
the SABER and X-shooter data sets did not allow us to find convincing
trends.

In agreement with \citet{NOL15}, non-LTE effects are critical for
absolute mesopause temperature estimates from \chem{OH}
$T_{\text{rot}}$ and comparisons of bands with different
$v^{\prime}$. Dynamical studies based on $T_{\text{rot}}$ derived from
\chem{OH} bands with high $v^{\prime}$ could also be significantly
affected. In this respect, only low $v^{\prime}$ appear to be
sufficiently safe.


\begin{acknowledgements}
  This project made use of the ESO Science Archive Facility. X-shooter
  spectra from different observing programmes of the period from
  October~2009 to March~2013 were analysed. Moreover, we thank the
  SABER team for providing the data products used in this paper. We
  thank the two anonymous referees for their detailed and very helpful
  comments. This publication is supported by the Austrian Science Fund
  (FWF). S. Noll and S. Unterguggenberger receive funding from the FWF
  project P26130. W. Kausch is funded by the project IS538003
  (Hochschulraumstrukturmittel) provided by the Austrian Ministry for
  Research (bmwfw).
\end{acknowledgements}

\end{document}